\documentclass[a4paper,titlepage,12pt]{article}

\usepackage{arxiv}

\usepackage[utf8]{inputenc} 					
\usepackage[english]{babel}
\usepackage[T1]{fontenc}                    
\usepackage{xcolor}
\definecolor{darkred}{rgb}{0.55, 0.0, 0.0}
\definecolor{darkcerulean}{rgb}{0.03, 0.27, 0.49}
\usepackage[colorlinks]{hyperref}          
\hypersetup{urlcolor={darkred}, citecolor={darkcerulean}}
\usepackage{url}                   					
\usepackage{booktabs}         					
\usepackage{titlesec}
\titlelabel{\thetitle.\quad}

\usepackage{amssymb,amsmath,amsfonts} 
\usepackage{calrsfs}
\usepackage{siunitx}     

\usepackage{nicefrac}       
\usepackage{microtype}      
\usepackage{lipsum}		

\usepackage[numbers]{natbib}

\usepackage[inline]{enumitem}

\usepackage{graphicx}
\usepackage{caption}
\usepackage{subcaption}
\usepackage[font=footnotesize,labelfont=bf]{caption}
\usepackage[font=footnotesize,labelfont=bf]{subcaption}

\usepackage{pgfplots}
\usepackage{pgfplotstable}
\pgfplotsset{compat=newest}

\usepgfplotslibrary{polar}

\usetikzlibrary{}
\usepgfplotslibrary{fillbetween}
\PassOptionsToPackage{gray}{xcolor}
\definecolor{orange}{gray}{0.4}

\usepackage{lscape}
\usepackage{tikz}
\usepackage{tikzscale}

\usetikzlibrary{positioning, shapes.geometric, arrows, fadings, arrows.meta, chains, positioning, spy, decorations.fractals, shapes.misc, calc, patterns, decorations.pathmorphing, decorations.markings, shapes, shadows, backgrounds}
\usepackage{csvsimple}

\usepackage{color}
\definecolor{red}{rgb}{0.5, 0.0, 0.13}

\title{Simplex space-time meshes in thermally coupled two-phase flow simulations of mold filling}

\date{} 					

\author{
Violeta~Karyofylli\thanks{Corresponding author\newline \textit{Email addresses:} \href{mailto:karyofylli@cats.rwth-aachen.de}{karyofylli@cats.rwth-aachen.de} (Violeta Karyofylli), \href{mailto:wendling@cats.rwth-aachen.de}{wendling@cats.rwth-aachen.de} (Lo\"ic Wendling), \href{mailto:make@cats.rwth-aachen.de}{make@cats.rwth-aachen.de} (Michel Make), \href{mailto:hosters@cats.rwth-aachen.de}{hosters@cats.rwth-aachen.de} (Norbert Hosters), \href{mailto:behr@cats.rwth-aachen.de}{behr@cats.rwth-aachen.de} (Marek Behr)  \newline
\newline \textit{NOTICE:} This is the author’s version of a work that was submitted for publication in \textit{Computers \& Fluids}. Changes resulting from the publishing process, such as editing, corrections, structural formatting, and other quality control mechanisms may not be reflected in this document. Changes may have been made to this work since it was submitted for publication.}\ , Lo\"ic~Wendling, Michel~Make, Norbert~Hosters, Marek~Behr}

\begin{document}
\maketitle

\begin{abstract}
The quality of plastic parts produced through injection molding depends on many factors. Especially during the filling stage, defects such as weld lines, burrs, or insufficient filling can occur. Numerical methods need to be employed to improve product quality by means of predicting and simulating the injection molding process. In the current work, a highly viscous incompressible non-isothermal two-phase flow is simulated, which takes place during the cavity filling. The injected melt exhibits a shear-thinning behavior, which is described by the Carreau-WLF model. Besides that, a novel discretization method is used in the context of 4D simplex space-time grids \cite{behr2008simplex}. This method allows for local temporal refinement in the vicinity of, e.g., the evolving front of the melt \cite{karyofylli2018simplex}. Utilizing such an adaptive refinement can lead to locally improved numerical accuracy while maintaining the highest possible computational efficiency in the remaining of the domain. For demonstration purposes, a set of 2D and 3D benchmark cases, that involve the filling of various cavities with a distributor, are presented.
\end{abstract}


\section{Introductions}
The thorough understanding of the filling stage in injection molding has been the subject of many experimental and numerical studies since it takes place in a wide variety of manufacturing procedures \cite{kim2004developments}. During this process, a highly viscous molten material is guided into a cavity and replaces a quiescent gas present inside the mold, meaning that several physical aspects need to be taken into consideration (e.g., the interplay of the two phases due to the discontinuities in their properties, the surface tension, the heat transfer, as well as the wetting, and shear-thinning effects) \cite{elgeti2012numerical}. A similar range of phenomena governs other production processes, such as die casting \cite{Siegbert2015}. Consequently, numerical methods and material models that are capable of modeling mold filing processes accurately and efficiently are inherently complex. 

In this work, a new method is presented, taking advantage of a 4D fully unstructured finite element (FE) formulation \cite{behr2008simplex}, founded on simplex space-time discretization. This method is suitable for interface capturing based on the level-set method and designed for two-phase incompressible flows in the context of mold filling. It allows for temporal refinement near areas of interest within the computational domain  \cite{karyofylli2018simplex}. With this method, the local computational accuracy is increased while maintaining, as much as possible, the computational efficiency. The current state of the art shows an increasing interest in 4D simplex space-time discretizations, where the time is considered as the fourth dimension. Noteworthy reasons for exploring unstructured space-time discretizations are the space-time mesh adaptivity, utilization of parallelism in 4D, and handling of topological changes. The following references are exploiting some of the advantages as mentioned above. In \cite{gopalakrishnan2018spacetime}, the discontinuous Petrov Galerkin (DPG) method is used for the solution of a transient acoustic wave system of equations in arbitrary dimensions. In addition, 4D FE discretizations are studied in \cite{VORONIN2018863} for the heat equation, the scalar conservation law and the wave equation, which are reformulated as constrained first-order system least-squares (CFOSLS). Regarding the linear algebra aspects, algebraic multigrid (AMG) preconditioned GMRES methods are investigated in \cite{steinbach2018comparison} for solving linear systems which arise from a space-time FE discretization of the heat equation in 3D and 4D space-time domains.

In the following sections of this publication, the governing equations, the material model and the corresponding solution techniques are presented, followed by a discussion of simplex space-time meshes when used for adaptive temporal refinement. Additionally, a set of numerical examples highlight the performance of the presented methods. Finally, some concluding remarks are given. 

\section{Theoretical background of mold filling} \label{Theoretical background of mold filling}

\subsection{Governing equations} \label{Governing equations}
During the filling stage of injection molding, a highly viscous incompressible non-isothermal two-phase flow takes place. Such flows are governed by the transient incompressible Navier-Stokes equations, coupled with the heat equation. 

The computational domain \(\Omega\), on which the governing equations need to be solved, is a subset of \(\mathbb{R}^{n_{sd}}\) and contains two immiscible fluids represented by the subdomains \(\Omega_1(t)\) and \(\Omega_2(t)\). Here, \(\Omega_1(t)\cup\Omega_2(t) = \Omega\). The boundary of \(\Omega\) is given by \(\Gamma = \partial{\Omega}\), while the interface between the two immiscible phases, i.e., melt and air, is defined as \(\Gamma_{\mathit{int}}(t)=\partial{\Omega_1(t)}\cap\partial{\Omega_2(t)}\).

At each instant \(t \in (0,t_{total}]\), the velocity, \(\mathbf{u}(\mathbf{x}, t )\), the pressure, \(p(\mathbf{x}, t )\), and the temperature, \(T(\mathbf{x}, t )\), in each phase \(i\) in subdomain \(\Omega_i(t)\) are governed by the following
equations:
\begin{align}
\rho_i\left(\frac{\partial \mathbf{u}}{\partial t}+\mathbf{u}\cdot\nabla\mathbf{u}-\mathbf{f}\right) -\nabla\cdot\pmb{\sigma}_i= \mathbf{0},  \label{Navier_Stokes} \\
\nabla\cdot\mathbf{u}= 0,  \label{Continuity_eq} \\
\pmb{\sigma}_i(\mathbf{u},p) = -p\mathbf{I}+2(\mu_{\mathit{eff}})_i \ \pmb{\varepsilon}(\mathbf{u}), \label{Stress_tensor} \\
\pmb{\varepsilon}(\mathbf{u})= \frac{1}{2}(\nabla\mathbf{u}+(\nabla\mathbf{u})^T ), \label{Strain_tensor} \\
\rho_{i} {\left( C_p \right)}_{i} (\frac{\partial T}{\partial t}+\mathbf{u}\cdot\nabla T) = k_{i} \bold{\nabla}^2 T + \Phi_{i}. \label{Heat_eq} 
\end{align}
where \(\rho_i\) is the density, \((\mu_{\mathit{eff}})_i\) is the dynamic effective viscosity, \(\pmb{\sigma}_i\) denotes the stress tensor, \(\pmb{\varepsilon}(\mathbf{u})\) stands for the rate-of-strain tensor and \({\left( C_p \right)}_{i}\) implies the isobaric heat capacity, which is assumed to be constant. Moreover, \(k_{i}\) is the thermal conductivity. Note that \((\mu_{\mathit{eff}})_i\) is a function of the shear rate, defined as \(G_f = \sqrt{2 \ \pmb{\varepsilon}(\mathbf{u}):\pmb{\varepsilon}(\mathbf{u})}\) for viscous incompressible fluids \cite{pauli2016stabilized}. The indexing \(i =1, 2\) represents the current phase. Finally, the viscous dissipation \(\Phi_{i}\) is defined, as follows:
\begin{equation} \label{ViscousDissipationFunction}
    \Phi_{i} = (\mu_{\mathit{eff}})_i  (\nabla\mathbf{u}+(\nabla\mathbf{u})^T ) : (\nabla\mathbf{u}+(\nabla\mathbf{u})^T ).
\end{equation}

The traction acting on the phase interface \(\Gamma_{\mathit{int}}(t)\) depends on the surface tension \(\gamma(\mathbf{x})\), and the interface curvature \(\kappa\). This leads to the following interface conditions \( \forall \ t \in (0,t_{total}]\):
\begin{align} \label{Surface_tension}
\mathbf{n}\cdot [ \pmb{\sigma}]_{\Gamma_{\mathit{int}}(t)} &=\kappa\gamma(\mathbf{x})\mathbf{n}, \\
\mathbf{t}\cdot [ \pmb{\sigma}]_{\Gamma_{\mathit{int}}(t)} &=\mathbf{\nabla}^{\Gamma_{\mathit{int}}(t)} \gamma(\mathbf{x}) \ \quad\mathrm{and} \label{eq:tangent-stress}\\
[\mathbf{u}]_{\Gamma_{\mathit{int}}(t)}&=\mathbf{0}.
\end{align}
where $[\mathbf\bullet]$ is the notation of the usual jump operator across \(\Gamma_{\mathit{int}}(t)\). Here, \(\mathbf{n}\) and \(\mathbf{t}\) denote the outward unit normal and tangent vector on \(\Gamma_{\mathit{int}}(t)\), respectively. The connection operator on \(\Gamma_{\mathit{int}}(t)\) is represented by \(\mathbf{\nabla}^{\Gamma_{\mathit{int}}(t)}\) in Equation \eqref{eq:tangent-stress}.  Although the surface tension \(\gamma(\mathbf{x})\) depends on a local surfactant concentration and temperature, as stated in \cite{seric2018direct}, in the current work \(\gamma(\mathbf{x})\) is assumed to be constant. Finally, compatibility between the two phases is enforced by requiring the velocities to be continuous across the interface. For the energy equation, the effects of surface tension, as described in \cite{blanchette2009energy}, are neglected.
 
For capturing the evolution of the interface between the two phases, as presented in \cite{osher1988fronts}, the level-set transport equation:
\begin{equation} \label{Level-Set}
\frac{\partial\phi}{\partial t}+\mathbf{u}\cdot\nabla\phi = 0 \quad  \mathrm{in} \quad  \Omega_i(t), \quad  \forall \ t \in (0,t_{total}] ,
\end{equation}
is solved. Here, \(\phi\) is a signed-distance function and \(\mathbf{u}\) is the velocity field obtained from the Navier-Stokes equations \eqref{Navier_Stokes} -- \eqref{Continuity_eq}.

After defining the level-set function, the boundary conditions regarding the velocity field on the walls \(\Gamma\) of \(\Omega\) needs to be described. Instead of using no-slip boundary conditions, which do not allow for the desired wetting during the filling stage of injection molding, the Navier-slip boundary condition is applied to the solid walls. It is a Robin-type boundary condition, as described in \cite{Behr2004}:
\begin{align}
\mathbf{n}_s \cdot \mathbf{u} &= 0, \label{eq:NavierSlipEquations1}\\
\mathbf{t}_s \cdot \pmb{\sigma} ( \mathbf{u}, p) \cdot \mathbf{u} &= \beta \mathbf{t}_s \cdot \mathbf{u}, \label{eq:NavierSlipEquations2}\\
\mathbf{b}_s \cdot \pmb{\sigma} ( \mathbf{u}, p) \cdot \mathbf{u} &= \beta \mathbf{b}_s \cdot \mathbf{u},
\label{eq:NavierSlipEquations3}
\end{align}
with $\mathbf{n}_s$ the normal, $\mathbf{t}_s$ the tangent, $\mathbf{b}_s$ the bi-tangent vector at the boundary and $\beta$ being the Navier-slip coefficient. Following the example of \cite{Sandia}, this coefficient is varying as follows:
\begin{equation}
\beta=\beta_{\infty} \cdot \delta \left( \phi \right) + \beta_0, \quad 
\delta \left( \phi \right) = 
    \begin{cases}
        \begin{aligned}
             0,&                   \quad   \quad  \quad   |\phi| \leq a& \\ 
             \frac{|\phi|}{a} -1,& \quad  a < |\phi| < 2a&   \\
             1,&                   \quad  \quad   \quad   |\phi| \geq 2a&
        \end{aligned}
   \end{cases}
\label{wetting}
\end{equation}
with $\beta_0$ being the wetting coefficient at the contact line, $\beta_{\infty}$ the far-field wetting coefficient, and $a$ the characteristic length scale around the interface. 

\subsection{Non-Newtonian fluids} \label{Non-Newtonian fluids}
In contrast to Newtonian fluids, the stresses in a non-Newtonian fluid are not linearly proportional to the local rate-of-strain. In the numerical results presented in this work, the molten material is assumed to be non-Newtonian. More specifically, a PZT ceramic paste (paste of lead zirconate titanate) is used and shows shear-thinning behavior. Therefore, the viscosity must be related to the shear rate through an appropriate model.

Following the example of \cite{Sandia}, the Carreau-Yasuda model has been used for the melts considered here:
\begin{equation}
    (\mu_{\mathit{eff}})_{\mathit{PZT}} = \mu_{\infty} + \left( \mu_0 - \mu_{\infty} \right) \left( 1 + \left(\lambda G_f \right) ^{a} \right) ^{\frac{\left( n - 1 \right)}{a}},
    \label{eq:Carreau}
\end{equation}
where $\mu_0$ is the viscosity at zero shear rate, $\mu_{\infty}$ the viscosity at infinite shear rate, $\lambda$ the relaxation time, $n$ the power-law index and $a$ the temperature sensitivity scaling factor. As stated in \cite{Sandia}, the Carreau-Yasuda material model is well-behaved at the zero shear-rate limit, in contrast to the power-law models. However, values for the low and high shear-rate viscosities must be defined.

Since the model depends on the fluid temperature, the equation needs to be corrected by a temperature shift factor $a_T$, as stated in \cite{osswald}. Thus, the corrected Carreau-Yasuda model, as specified by \cite{Sandia}, has the form:
\begin{equation}
    (\mu_{\mathit{eff}})_{\mathit{PZT}} = a_T \left[ \mu_{\infty} + \left( \mu_0 - \mu_{\infty} \right) \left( 1 + \left( a_T \lambda G_f \right) ^{a}\right) ^{\frac{\left( n - 1 \right)}{a}} \right].
    \label{eq:ModifiedCarreau}
\end{equation}
Although several different formulations for $a_T$ exist, the Williams-Landel-Ferry (WLF) formulation is used here:
\begin{equation}
    a_T = exp \left[ \frac{c_1 \left( T_{ref} - T \right)}{c_2 + T - T_{ref}}  \right],
    \label{eq:WLFShiftFactor}
\end{equation}
with $c_1$, $c_2$ and $T_{ref}$ being the WLF parameters and the reference temperature, respectively. They are empirically obtained via regression to experimental data. The WLF shift factor allows for the estimation of the viscosity curves for different temperatures than those at which the molten plastic was tested.

\begin{figure}[!htb]
    \centering
    \includegraphics[width=0.75\textwidth]{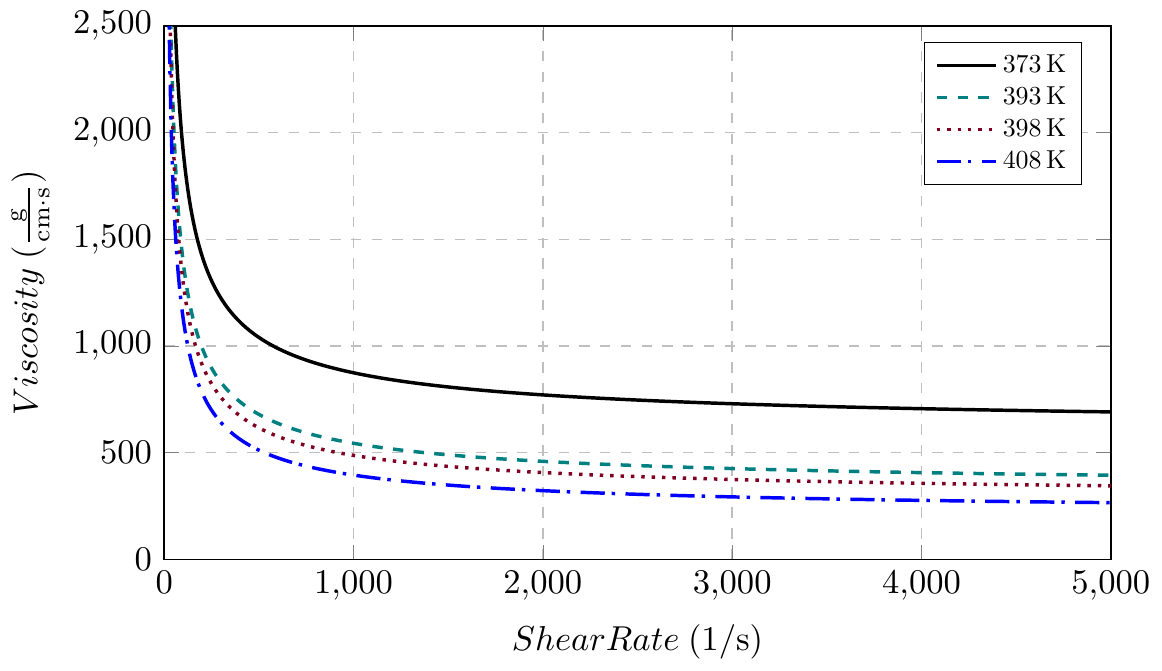}
    \caption{Viscosity according to Carreau-Yasuda-WLF model for four different temperatures ($\SI{373}{K}$, $\SI{393}{K}$, $\SI{398}{K}$ and $\SI{408}{K}$).}
    \label{fig:ViscosityCarreauWlfComparison}
\end{figure}

\begin{table}[ht]
\caption{Non-isothermal Carreau-Yasuda-WLF parameters}
\centering
\begin{tabular}{c c}
\hline\hline
Carreau-Yasuda-WLF parameter                 & Value                      \\ [0.5ex] 
\hline
low shear-rate viscosity, $\mu_0$ & \SI{50000}{\big[\frac{g}{cm \cdot s}\big]} \\
high shear-rate viscosity, $\mu_\infty$    & \SI{350}{\big[\frac{g}{cm \cdot s}\big]}      \\
Power-law coefficient, n & \SI{0.32}{[-]} \\
Time constant, $\lambda$ & \SI{2.67}{[s]} \\
Scale factor, $a$ & \SI{1.915}{[-]} \\
WLF parameter, $c_1$ & \SI{50}{[-]} \\ 
WLF parameter, $c_2$ & \SI{1600}{[-]} \\ 
WLF reference temperature, $T_{ref}$ & \SI{390}{[K]} \\ [1ex]
\hline
\end{tabular}
\label{table:Carreau-Yasuda-WLF}
\end{table}
The parameter values for the Carreau-Yasuda-WLF model are presented in Table \ref{table:Carreau-Yasuda-WLF} and are those used later in the numerical examples. Correcting the Carreau-Yasuda model with the WLF formulation leads to the shear rate and temperature dependency of the viscosity as presented in Figure \ref{fig:ViscosityCarreauWlfComparison}.

Last but not least, concerning our material model, we need to draw the attention to the thermal conductivity coefficient of the PZT ceramic paste \(k_{\mathit{PZT}}\), which is coupled to the dynamic viscosity through the Prandtl number as follows:
\begin{equation} \label{ThermalCond}
    k_{\mathit{PZT}} = \frac{C_p \mu_\infty}{Pr}.
\end{equation}
 This leads to a better convergence of the heat equation. The Prandtl number is computed based on the parameters, described in Tables \ref{table:Carreau-Yasuda-WLF} and \ref{table:MaterialProperties2D}:
\begin{equation} \label{PrandtlNumber}
    Pr = \frac{C_p \mu_\infty}{(k_0)_{\mathit{PZT}}}.
\end{equation}

\subsection{Solution technique} \label{Solution technique}
In order to discretize equations \eqref{Navier_Stokes}, \eqref{Continuity_eq}, \eqref{Heat_eq} and \eqref{Level-Set}, \(P1P1\) finite elements are used. The Galerkin/least-squares (GLS) stabilization method is applied for this purpose. In the GLS method, the stabilization term is a least-squares form of the original differential equation, weighted element by element \cite{Donea}. For the creation of the finite element function spaces, which are used for the space-time discretization, the time interval is divided into subintervals \(I_n = (t_n, t_{n+1})\), with \(t_n\) and \(t_{n+1}\) being two consecutive time levels which belong to \(0 = t_0 < t_1 < \cdots < t_N = T\). If \(\Omega_n = \Omega({t_n})\), the space-time slab  \(Q_n\) is defined as the region confined by the surfaces \(\Omega_n, \Omega_{n+1}\) and the space-time boundary, described by the spatial boundary \(\Gamma(t)\) evolved over \(I_n\), referred to as \(P_n\), and contains the advancing space-time front, \((P_{\mathit{int}})_n\). The finite element interpolation and weighting function spaces which are set for the variables of the equation system (velocity, pressure, level-set function, and temperature) are:
\begin{align}
(\mathbf{\mathcal{S}}_\mathbf{u}^h)_n &= \{\mathbf{u}^h \ | \ \mathbf{u}^h \in [H^{1h}(Q_n)]^{n_{sd}}, \ \mathbf{u}^h \dot{=} \mathbf{g}^h \quad \mathrm{on} \quad (P_n)_\mathbf{u}\}, \label{Interpolation_fs_vel} \\
(\mathcal{V}_\mathbf{u}^h)_n &= \{\mathbf{w}^h \ | \ \mathbf{w}^h \in [H^{1h}(Q_n)]^{n_{sd}}, \ \mathbf{w}^h \dot{=} \mathbf{0} \quad \mathrm{on} \quad (P_n)_\mathbf{u}\}, \label{Weighting_fs_vel} \\
(\mathcal{S}_p^h)_n &= (\mathcal{V}_p^h)_n = \{p^h \ | \ p^h \in L^{2h}(Q_n)\} , \label{Interpolation_Weighting_fs_pres} \\
(\mathbf{\mathcal{S}}_T^h)_n &= \{T^h \ | \ T^h \in H^{1h}(Q_n), \ T^h \dot{=} \hat{T}^h \quad \mathrm{on} \quad (P_n)_{T}\} \quad \label{Interpolation_fs_T} \\
(\mathcal{V}_T^h)_n &= \{v^h \ | \ v^h \in H^{1h}(Q_n), \ v^h \dot{=} 0 \quad \mathrm{on} \quad (P_n)_{T}\}, \label{Weighting_fs_T} \\
(\mathbf{\mathcal{S}}_\phi^h)_n &= \{\phi^h \ | \ \phi^h \in H^{1h}(Q_n), \ \phi^h \dot{=} \hat{\phi}^h \quad \mathrm{on} \quad (P_n)_{\phi}\} \ \mathrm{and} \label{Interpolation_fs_fi} \\
(\mathcal{V}_\phi^h)_n &= \{\psi^h \ | \ \psi^h \in H^{1h}(Q_n), \ \psi^h \dot{=} 0 \quad \mathrm{on} \quad (P_n)_{\phi}\}. \label{Weighting_fs_fi}
\end{align}
for every space-time slab. Here, \(H^{1h} \subset H^1\) is a finite dimensional Sobolev space consisting of functions which are square-integrable in \(Q_n\) and have square-integrable first derivatives. The trial function spaces for the velocity, level-set function and temperature, denoted by \((\mathbf{\mathcal{S}}_\bullet^h)_n\), must additionally satisfy the Dirichlet boundary conditions. Similar test function spaces \((\mathcal{V}_\bullet^h)_n\) are chosen, but it is required that the test functions
vanish on the Dirichlet boundary \(P_n\). The requirements for the pressure trial function space are less restrictive, as no derivatives of the pressure appear and no explicit pressure boundary conditions exist. Therefore, the pressure trial and test functions are chosen
from \(L^{2h} \subset L^2\), being the finite dimensional Hilbert space of square-integrable functions. The interpolation functions in the elements constitute first-order polynomials, which are continuous in space, but discontinuous in time for every region of the domain.

The Navier-Stokes equations \eqref{Navier_Stokes} and \eqref{Continuity_eq}, the heat equation \eqref{Heat_eq} and the level-set transport equation \eqref{Level-Set} have the following space-time discretized forms, respectively, after stabilization: 

Given \((\mathbf{u}^h)_n^-\), find \(\mathbf{u}^h \in (\mathcal{S}_\mathbf{u}^h)_n\) and \(p^h \in (\mathcal{S}_p^h)_n\) such that \(\forall \mathbf{w}^h \in (\mathcal{V}_\mathbf{u}^h)_n, \forall q^h \in (\mathcal{V}_p^h)_n\):
\begin{align}\label{Variational_Navier-Stokes}
\int_{Q_n}\mathbf{w}^h&\cdot\rho_i\left(\frac{\partial \mathbf{u}^h}{\partial t}+\mathbf{u}^h\cdot\nabla\mathbf{u}^h-\mathbf{f}\right) \ dQ + \int_{Q_n}\pmb{\varepsilon}(\mathbf{w}^h):\pmb{\sigma}_i(\mathbf{u}^h,p^h) \ dQ \nonumber \\
&+ \int_{Q_n}q^h\nabla\cdot\mathbf{u}^h\ dQ + \int_{\Omega_n}(\mathbf{w}^h)^+_n\cdot\rho_i((\mathbf{u}^h)^+_n-(\mathbf{u}^h)^-_n) \ d\Omega \nonumber \\
&+\sum_{e=1}^{(n_{el})_n} \int_{Q_n^e}\tau_{MOM}\frac{1}{\rho_i} \left[\rho_i \left(\frac{\partial \mathbf{w}^h}{\partial t}+\mathbf{u}^h\cdot\nabla \mathbf{w}^h \right)-\nabla\cdot\pmb{\sigma}_i(\mathbf{w}^h, q^h) \right] \nonumber \\
&\cdot \left[\rho_i \left(\frac{\partial \mathbf{u}^h}{\partial t}+\mathbf{u}^h\cdot\nabla \mathbf{u}^h -\mathbf{f}\right)-\nabla\cdot\pmb{\sigma}_i(\mathbf{u}^h, p^h) \right] \ dQ \nonumber\\
&+\sum_{e=1}^{(n_{el})_n}\int_{Q_n^e}\tau_{CONT}\nabla\cdot\mathbf{w}^h\rho_i\nabla\cdot\mathbf{u}^h\ dQ \nonumber \\
&= \int_{(P_{\mathit{int}})_n}\mathbf{w}^h\cdot\gamma\kappa\mathbf{n} \ dP + \int_{(P_n)_{h}}\mathbf{w}^h\cdot\mathbf{h}^h \ dP . 
\end{align}

Given \((T^h)_n^-\), find \(T^h \in (\mathcal{S}_T^h)_n\) such that \(\forall v^h \in (\mathcal{V}_T^h)_n\):
\begin{align} \label{Variational_Heat_Equation}
\int_{Q_n}v^h&\left(\frac{\partial T^h}{\partial t} +\bold{u}^h\cdot\nabla T^h - \frac{k_{i}}{\rho_{i} {\left( c_v \right)}_{i}} \bold{\nabla}^2 T^h - \Phi_{i}(\bold{u}^h)\right) \ dQ \nonumber \\
&+ \int_{\Omega_n}(v^h)^+_n((T^h)^+_n-(T^h)^-_n) \ d\Omega \nonumber \\
&+ \sum_{e=1}^{n_{el}} \int_{Q_n^e}\tau_{TEMP}\left[\frac{\partial v^h}{\partial t}+\bold{u}^h\cdot\nabla v^h- \frac{k_{i}}{\rho_{i} {\left( c_v \right)}_{i}} \bold{\nabla}^2 v^h\right] \nonumber \\
&\left[\frac{\partial T^h}{\partial t} +\bold{u}^h\cdot\nabla T^h - \frac{k_{i}}{\rho_{i} {\left( c_v \right)}_{i}} \bold{\nabla}^2 T^h - \Phi_{i}(\bold{u}^h)\right] \ dQ = 0 . 
\end{align}

Given \((\phi^h)_n^-\), find \(\phi^h \in (\mathcal{S}_\phi^h)_n\) such that \(\forall \psi^h \in (\mathcal{V}_\phi^h)_n\):
\begin{align} \label{Variational_Level-Set}
\int_{Q_n}\psi^h&\left(\frac{\partial\phi^h}{\partial t} +\bold{u}^h\cdot\nabla\phi^h\right) \ dQ + \int_{\Omega_n}(\psi^h)^+_n((\phi^h)^+_n-(\phi^h)^-_n) \ d\Omega \nonumber \\
&+ \sum_{e=1}^{n_{el}} \int_{Q_n^e}\tau_{LEV}\left[\frac{\partial \psi^h}{\partial t}+\bold{u}^h\cdot\nabla \psi^h\right]\left[\frac{\partial\phi^h}{\partial t}+\bold{u}^h\cdot\nabla\phi^h\right] \ dQ = 0 . 
\end{align}

In the above equations, the notation below is applied:
\begin{align} \label{Notation}
(\mathbf\bullet^h)^\pm_n&=\lim_{\varepsilon\to0}\mathbf{\bullet}(t_n\pm\varepsilon), \nonumber \\
\int_{Q_n} \ldots dQ&=\int_{I_n}\int_{\Omega^h_t} \ldots d\Omega dt \quad \mathrm{and}  \\
\int_{P_n} \ldots dP&=\int_{I_n}\int_{\Gamma^h_t} \ldots d\Gamma dt . \nonumber
\end{align}
The problem is solved sequentially for each space-time slab, starting with \((\mathbf\bullet^h)^+_n=\mathbf\bullet_0\) at \(t_0\). We use a metric stabilization approach. Further information regarding this stabilization method and its parameters \(\tau_{MOM}\) and  \(\tau_{CONT}\) is given in \cite{2018arXiv181202070V} and \cite{pauli2017stabilized}, while the parameters \(\tau_{LEV}\) and \(\tau_{TEMP}\) are similarly computed. Moreover, the surface tension term in \eqref{Variational_Navier-Stokes} can be transformed through the Laplace-Beltrami technique, as suggested in \cite{hysing, sauerland, Elgeti2016}, into:
\begin{align} \label{Laplace_Beltrami}
\int_{(P_{\mathit{int}})_n}\mathbf{w}^h\cdot\gamma\kappa\mathbf{n} \ dP &= \int_{(P_{\mathit{int}})_n}\mathbf{w}^h\cdot\gamma\underline\Delta\mathbf{id}_{(P_{\mathit{int}})_n} \ dP \nonumber \\
 &= - \int_{(P_{\mathit{int}})_n}\gamma\underline\nabla\mathbf{id}_{(P_{\mathit{int}})_n}:\underline\nabla\mathbf{w}^h \ dP ,
\end{align}
with \(\underline\Delta\) being the Laplace-Beltrami operator, \(\underline\nabla\) the tangential gradient and \(\mathbf{id}\) the identity mapping on the space-time evolving interface \((P_{\mathit{int}})_n\).

The last term $\int_{(P_n)_{h}}\mathbf{w}^h\cdot\mathbf{h}^h \ dP$ in \eqref{Variational_Navier-Stokes} is of primary importance for applying the Navier-slip boundary condition to the cavity walls (as stated in \cite{karyofylli2017novel}), as it is a formulation to apply traction \(\mathbf{h}^h\) to a boundary of the domain. The term \(\mathbf{h}\) corresponds to the imposed traction, whereas the superscript \(h\) stands for discretized quantities.

\subsection{Coupling between Navier-Stokes, level-set and heat equations}
\begin{figure}[!htb]
    \centering
    \includegraphics[width=0.75\textwidth]{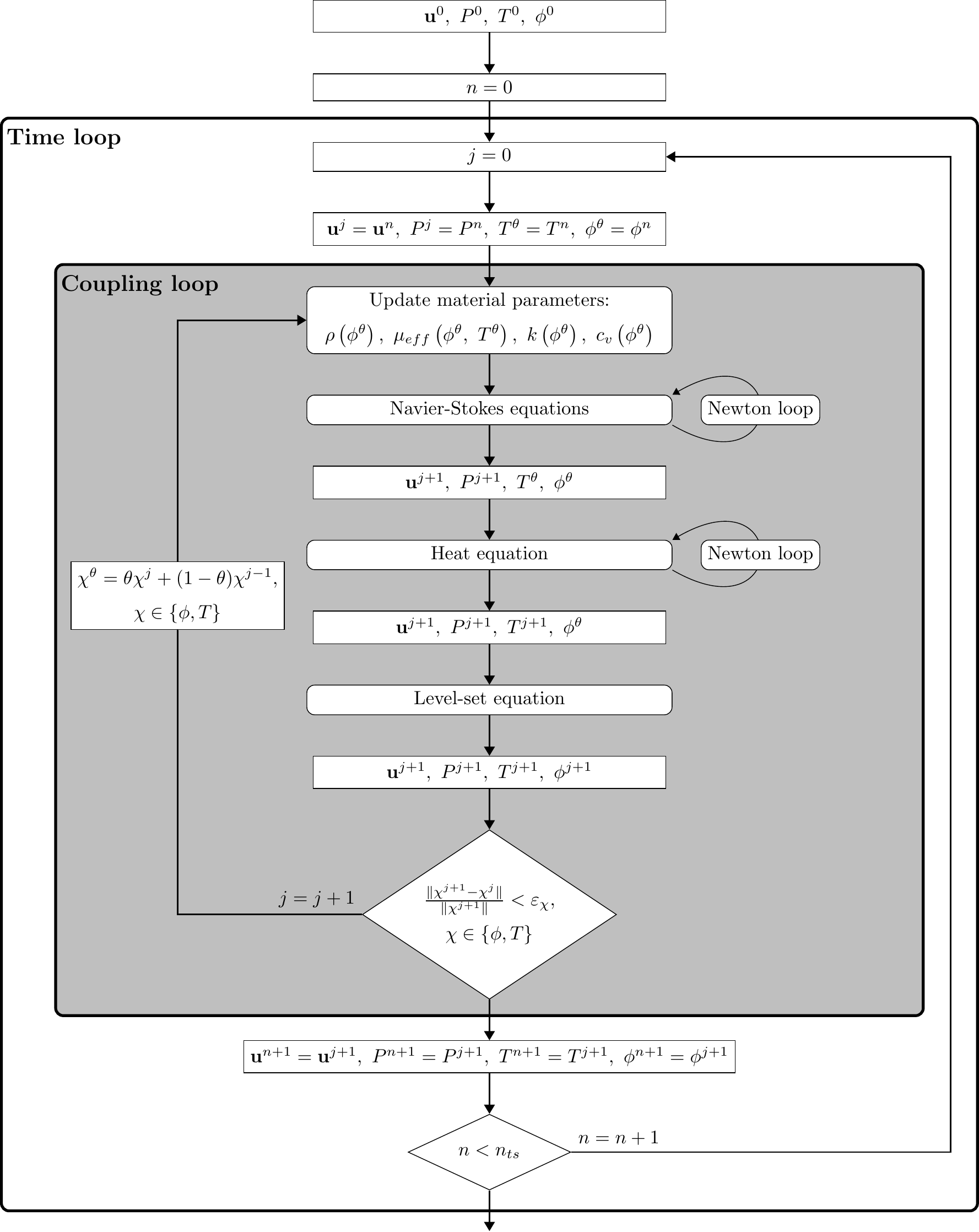}
    \caption{A strongly coupled partitioned approach for the Navier-Stokes, level-set and heat equations.}
    \label{fig: CouplingFlowChart}
\end{figure}

In this section, it is worth pointing out the mutual interaction between the fluid velocity, level-set, and temperature fields. On the one hand, the quantities that depend on the level-set field are the surface tension coefficient, the density, the viscosity, the isobaric heat capacity, and the thermal conductivity. On top of that, the viscosity of the melt is also a function of the temperature field, since the melt is considered to behave as a shear-thinning fluid (cf. Section \ref{Non-Newtonian fluids}). On the other hand, the level-set and temperature fields are advected by the fluid velocity. For this paper, a partitioned approach for the coupling has been chosen, meaning that the flow field is computed using a fixed melt-air interface. The temperature distribution is then calculated utilizing the already determined flow field and assuming a fixed melt-air front. Finally, the velocity field advects the melt-air interface. If this procedure is performed once per time step, then we refer to the coupling approach as weak. If the previously described process is repeated until all fields are in equilibrium and the desired convergence is reached, as described schematically in Figure \ref{fig: CouplingFlowChart}, then a strong coupling is obtained. The coupling is repeated at every time step, $n$, until the total number of time steps, $n_{ts}$, is reached. If a strongly coupled partitioned approach is used, the level-set, \(\phi^{\theta}\), and temperature field, \(T^{\theta}\), is a relaxation between the level-set and temperature field from the previous coupling iteration, \(\phi^{j-1}\) and \(T^{j-1}\), and those from the current iteration, \(\phi^j\) and \(T^j\), respectively.

On the one hand, a strongly coupled partitioned approach is more precise, providing more stable results in comparison to a weakly coupled partitioned approach. On the other hand, it requires more computation time, as stated in \cite{sauerland, reusken2017finite}.

\section{Simplex space-time finite elements}
An algorithm for the construction of simplex space-time meshes is proposed by \citep{behr2008simplex}. If $n_{sd}$ is the number of spatial dimensions, an $\left( n_{sd} + 1 \right)$-simplex is a space-time element, created after extruding the spatial simplex element in time and subdividing it into simplices, and contains $\left( n_{sd} + 2 \right)$ vertices. Furthermore, we need to point out that an $n_{sd}$-dimensional spatial simplex extruded in time can be subdivided into $n_{sd} + 1$ simplicial elements.

The simplex space-time elements allow for local temporal refinement, by adding nodes between the top and the bottom of the space-time slab. In the case of two-phase flow problems, an algorithm that applies temporal refinement in the vicinity of the evolving interface is described in \cite{karyofylli2018simplex}.

For setting the boundary conditions \eqref{eq:NavierSlipEquations1} -- \eqref{eq:NavierSlipEquations3}, the computation of the local normal--tangent--bi-tangent coordinate system at the solid boundaries of the mold must be performed, as described in Section \ref{Governing equations}. This computation is based on a generalized method for calculating the cross-product in higher dimensions, as stated in \cite{Lehrenfeld, neumuller2011refinement}.

\section{Numerical Results} \label{NumericalResults}

\subsection{2D cavity with a distributor}
As a first test case, we simulate the filling stage of a 2D geometry consisting of a pipe and a cavity, connected through a distributor. This benchmark case was also simulated in \cite{Sandia} with the software GOMA, developed by Sandia National Laboratories.
\begin{figure}[!h]
    \centering
    \includegraphics[width=0.75\textwidth]{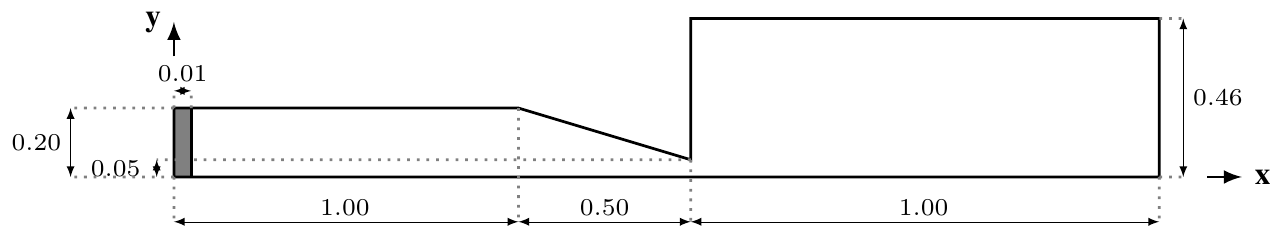}
    \caption{The computational domain of the two-dimensional cavity, as described in \cite{Sandia}. All dimensions are in $\SI{}{cm}$. The initial position of the molten material covers a distance of $\SI{0.01}{cm}$ inside the pipe (grey region).}
    \label{fig:2DSandiaGeometry}
\end{figure}

Figure \ref{fig:2DSandiaGeometry} illustrates the computational domain. The spatial discretization of the domain consists of \(\SI{27334}{}\) triangular elements. The time-slab size varies between \(\Delta t = \SI{0.001}{s} - \SI{0.005}{s}\). Navier-slip boundary condition is assumed for the velocity degrees of freedom on the walls, except for the inflow (leftmost vertical boundary) and outflow (rightmost vertical boundary). The molten PZT ceramic paste enters the cavity with a parabolic inflow velocity profile (\(u = 168 \frac{y}{0.2}\left(1 - \frac{y}{0.2}\right)\SI{}{\frac{cm}{s}}\)) and displaces the air, which is initially quiescent. Traction-free boundary condition is used at the outflow boundary. With respect to the temperature degree of freedom, homogeneous Neumann boundary conditions are imposed on the mold walls, except for the inflow one, where the injection temperature is equal to \(\SI{405}{K}\). Although we consider non-isothermal conditions, the phase-change effects are disregarded. The material properties correspond to those of \cite{Sandia} and are shown in Table \ref{table:MaterialProperties2D}. The gravitational acceleration is neglected, but the surface tension coefficient remains constant and equal to \(\gamma(\mathbf{x}) = \SI{42.4}{\frac{g}{s^2}}\).

\begin{table}[ht]
\caption{Material properties}
\centering
\begin{tabular}{c c}
\hline\hline
Material property                 & Value                      \\ [0.5ex] 
\hline
Density of paste, $\rho_{\mathit{PZT}}$ & \SI{4.5}{\big[\frac{g}{cm^3}\big]} \\
Viscosity of paste, $(\mu_{\mathit{eff}})_{\mathit{PZT}}$    & See Section \ref{Non-Newtonian fluids}      \\
Heat capacity of paste, ${\left( C_p \right)}_{\mathit{PZT}}$ & \SI{6200000}{\big[\frac{cm^2}{s^2 \cdot K}\big]} \\
Heat conductivity of paste, $(k_0)_{\mathit{PZT}}$ & \SI{61000}{\big[\frac{g \cdot cm}{K}\big]} \\
Density of air, $\rho_{air}$ & \SI{0.001}{\big[\frac{g}{cm^3}\big]} \\
Viscosity of air, $(\mu_{\mathit{eff}})_{air}$    & \SI{0.1}{\big[\frac{g}{cm \cdot s}\big]}      \\
Heat capacity of air, ${\left( C_p \right)}_{air}$ & \SI{10060000}{\big[\frac{cm^2}{s^2 \cdot K}\big]} \\
Heat conductivity of air, $k_{air}$ & \SI{2623}{\big[\frac{g \cdot cm}{K}\big]} \\ [1ex]
\hline
\end{tabular}
\label{table:MaterialProperties2D}
\end{table}

\begin{figure}[!h]
    \centering
    \includegraphics[width=1.0\textwidth]{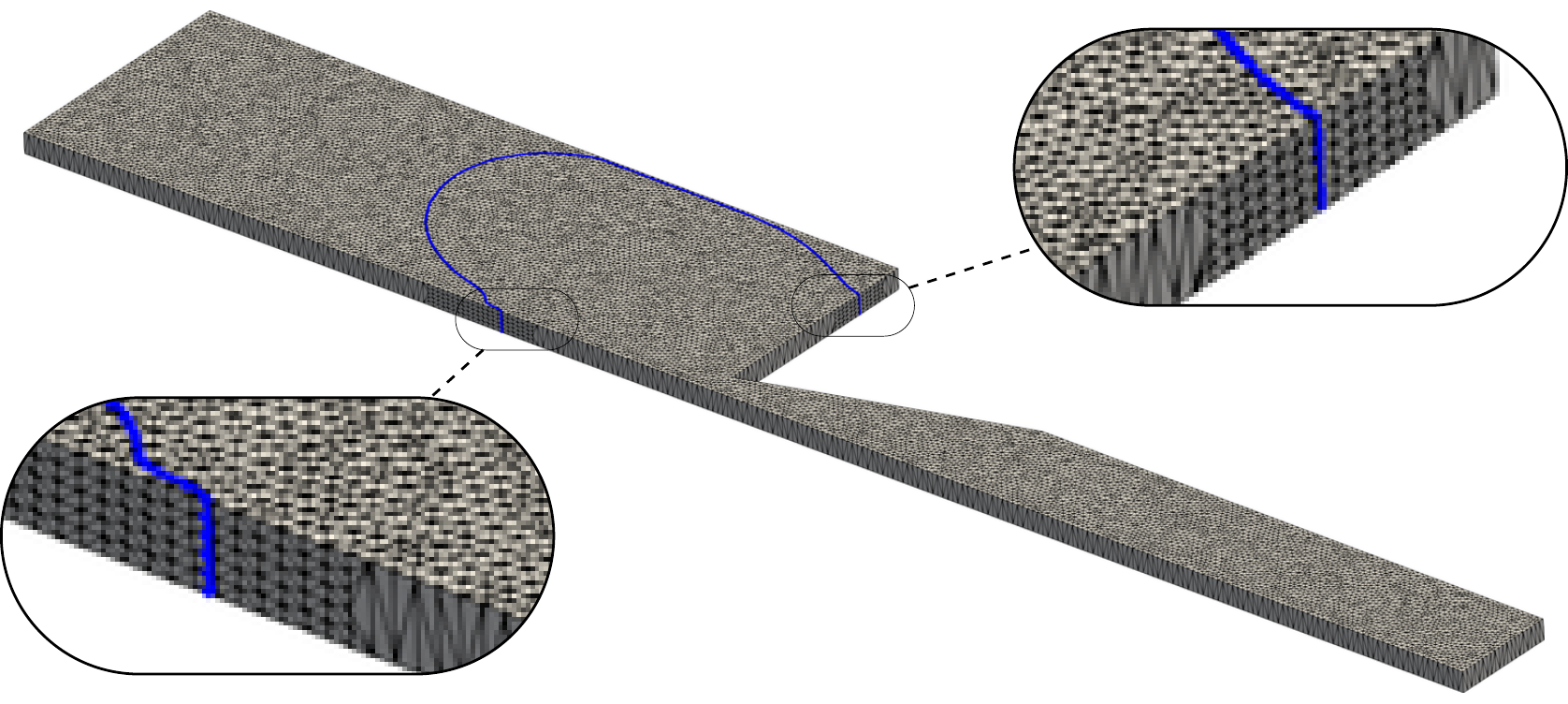}
    \caption{Adaptive temporal refinement is applied within an interval of $\SI{0.05}{cm}$ around the interface, by inserting four additional nodes in time direction. The thickness of the time-slab is scaled for visualization purposes.}
    \label{fig:TemporalRefinement}
\end{figure}

\begin{figure}[!htb]
    \centering
    \begin{subfigure}[t]{0.75\textwidth}
        \includegraphics[width=\textwidth]{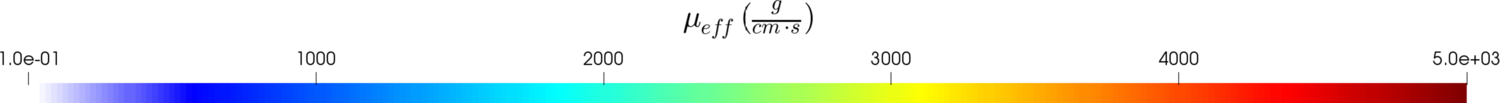}
        \label{fig:2DSandiaCavityVelocityRam}
    \end{subfigure}\\
    
    \begin{subfigure}[t]{0.3\textwidth}
        \reflectbox{\includegraphics[width=\textwidth]{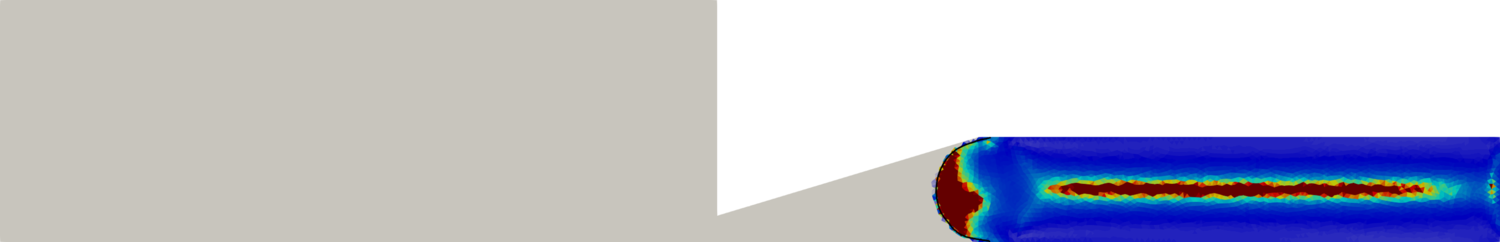}}
        \caption{FST (fine) at $t = \SI{0.0375}{s}$}
        \label{fig:2DSandiaCavityViscDt0_0001_15Nrec}
    \end{subfigure}
    ~ 
    \begin{subfigure}[t]{0.3\textwidth}
        \reflectbox{\includegraphics[width=\textwidth]{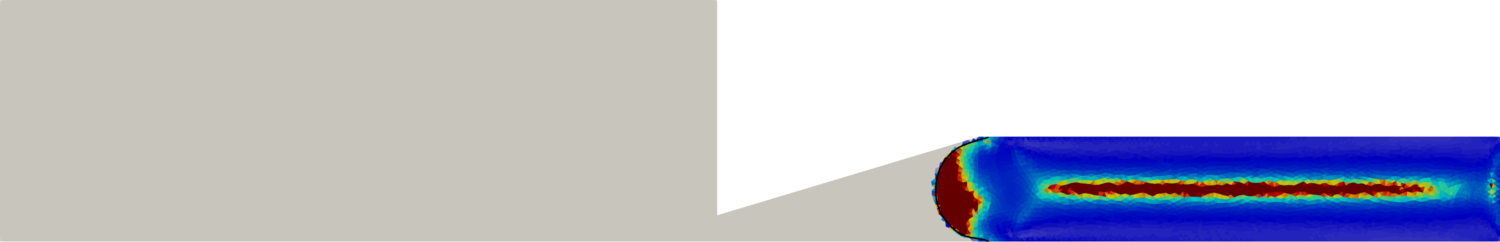}}
        \caption{SST at $t = \SI{0.0375}{s}$}
        \label{fig:2DSandiaCavity_ViscTimeRef15Nrec}
    \end{subfigure}
    ~ 
    \begin{subfigure}[t]{0.3\textwidth}
        \reflectbox{\includegraphics[width=\textwidth]{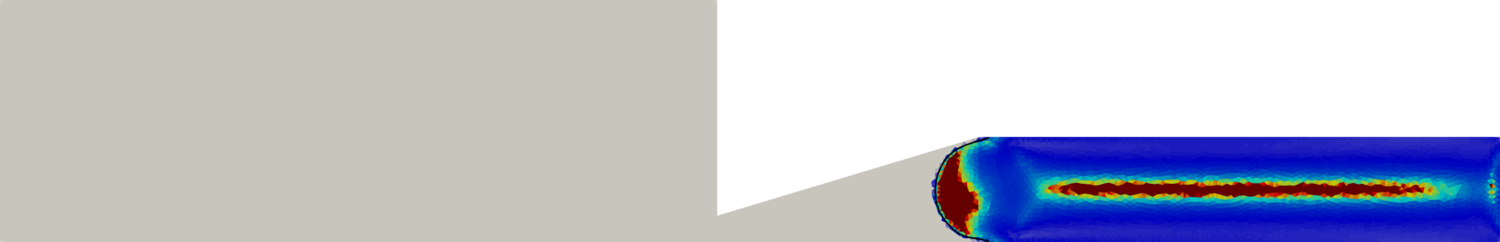}}
        \caption{FST (coarse) at $t = \SI{0.0375}{s}$}
        \label{fig:2DSandiaCavityViscDt0_0005_15Nrec}
    \end{subfigure}\\
    
    \begin{subfigure}[t]{0.3\textwidth}
        \reflectbox{\includegraphics[width=\textwidth]{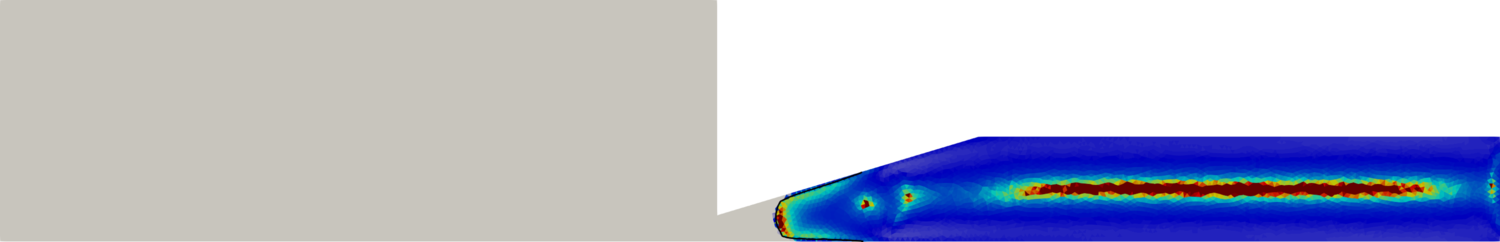}}
        \caption{FST (fine) at $t = \SI{0.0450}{s}$}
        \label{fig:2DSandiaCavityViscDt0_0001_18Nrec}
    \end{subfigure}
    ~ 
    \begin{subfigure}[t]{0.3\textwidth}
        \reflectbox{\includegraphics[width=\textwidth]{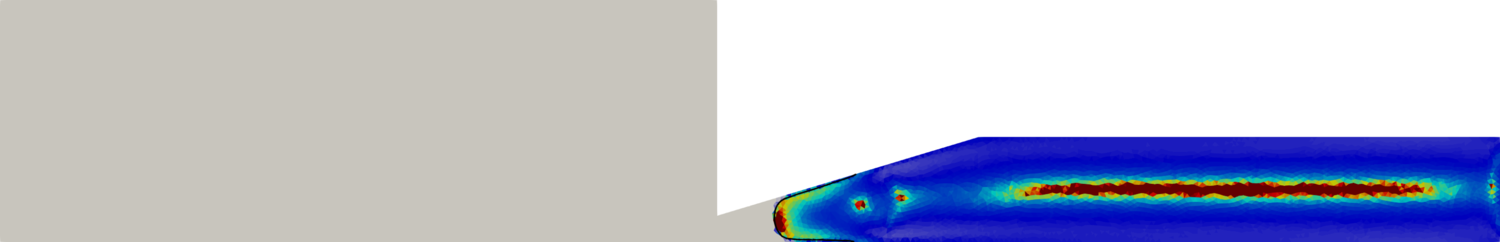}}
        \caption{SST at $t = \SI{0.0450}{s}$}
        \label{fig:2DSandiaCavity_ViscTimeRef18Nrec}
    \end{subfigure}
    ~ 
    \begin{subfigure}[t]{0.3\textwidth}
        \reflectbox{\includegraphics[width=\textwidth]{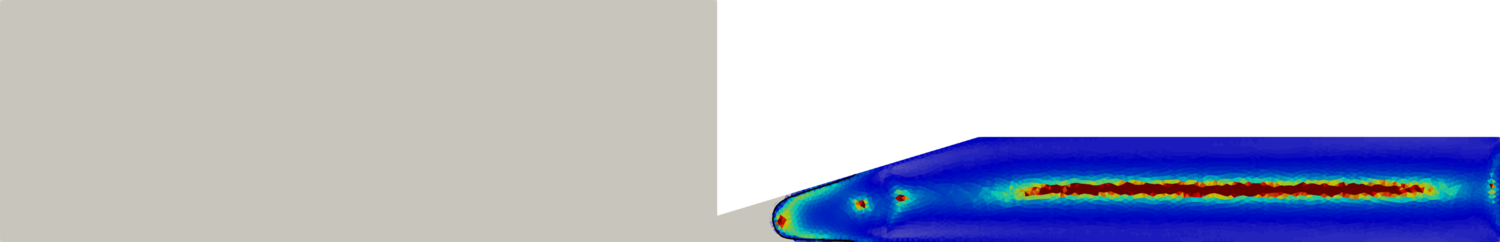}}
        \caption{FST (coarse) at $t = \SI{0.0450}{s}$}
        \label{fig:2DSandiaCavityViscDt0_0005_18Nrec}
    \end{subfigure}\\
    
     \begin{subfigure}[t]{0.3\textwidth}
        \reflectbox{\includegraphics[width=\textwidth]{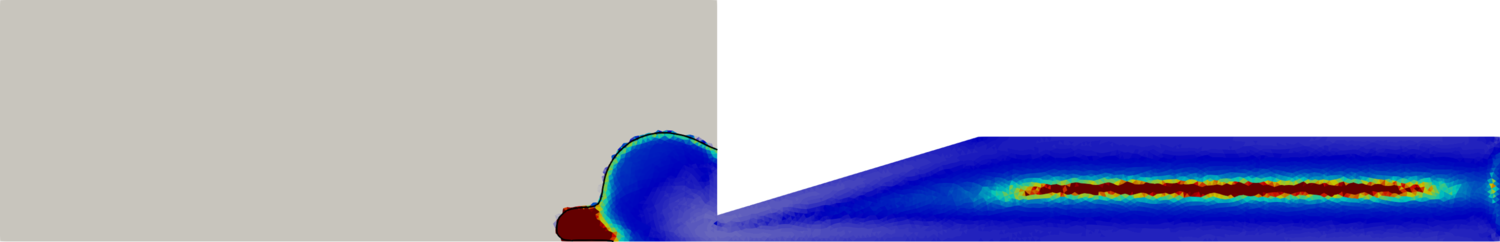}}
        \caption{FST (fine) at $t = \SI{0.0550}{s}$}
        \label{fig:2DSandiaCavityViscDt0_0001_22Nrec}
    \end{subfigure}
    ~ 
    \begin{subfigure}[t]{0.3\textwidth}
        \reflectbox{\includegraphics[width=\textwidth]{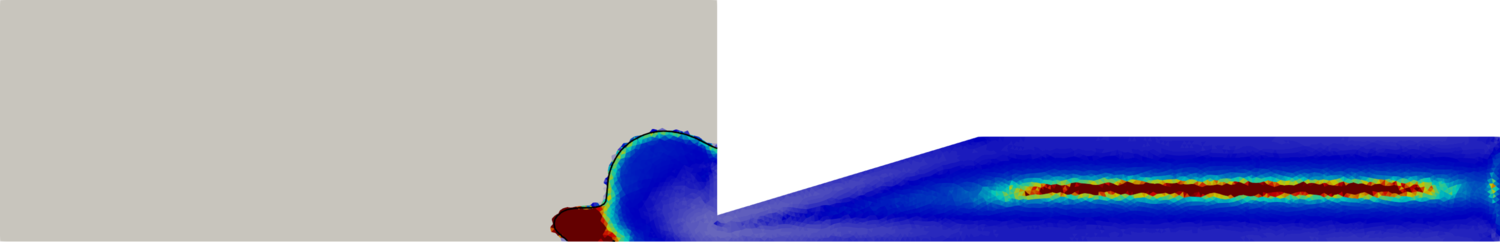}}
        \caption{SST at $t = \SI{0.0550}{s}$}
        \label{fig:2DSandiaCavity_ViscTimeRef22Nrec}
    \end{subfigure}
    ~ 
    \begin{subfigure}[t]{0.3\textwidth}
        \reflectbox{\includegraphics[width=\textwidth]{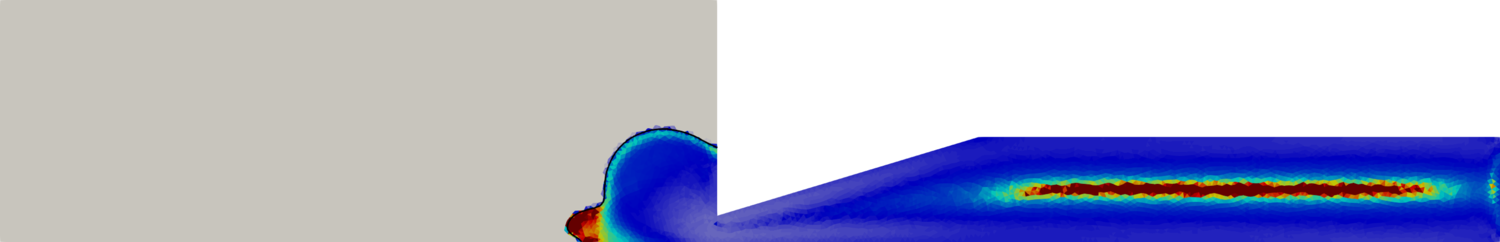}}
        \caption{FST (coarse) at $t = \SI{0.0550}{s}$}
        \label{fig:2DSandiaCavityViscDt0_0005_22Nrec}
    \end{subfigure}\\
    
    \begin{subfigure}[t]{0.3\textwidth}
        \reflectbox{\includegraphics[width=\textwidth]{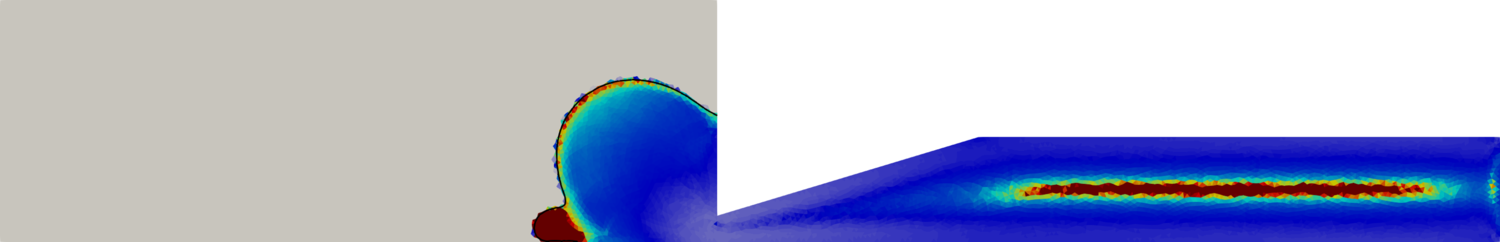}}
        \caption{FST (fine) at $t = \SI{0.0625}{s}$}
        \label{fig:2DSandiaCavityViscDt0_0001_25Nrec}
    \end{subfigure}
    ~ 
    \begin{subfigure}[t]{0.3\textwidth}
        \reflectbox{\includegraphics[width=\textwidth]{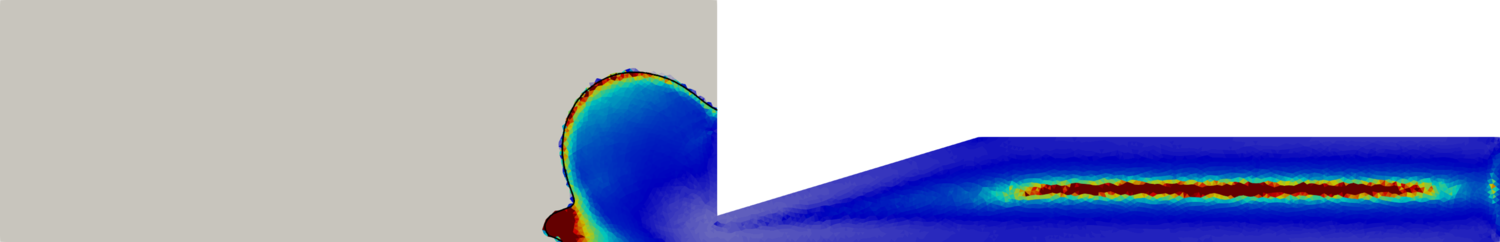}}
        \caption{SST at $t = \SI{0.0625}{s}$}
        \label{fig:2DSandiaCavity_ViscTimeRef25Nrec}
    \end{subfigure}
    ~ 
    \begin{subfigure}[t]{0.3\textwidth}
        \reflectbox{\includegraphics[width=\textwidth]{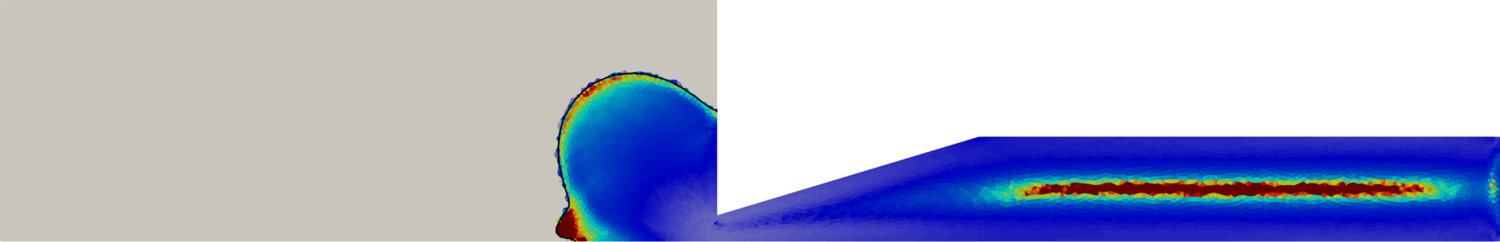}}
        \caption{FST (coarse) at $t = \SI{0.0625}{s}$}
        \label{fig:2DSandiaCavityViscDt0_0005_25Nrec}
    \end{subfigure}\\
    
    \begin{subfigure}[t]{0.3\textwidth}
        \reflectbox{\includegraphics[width=\textwidth]{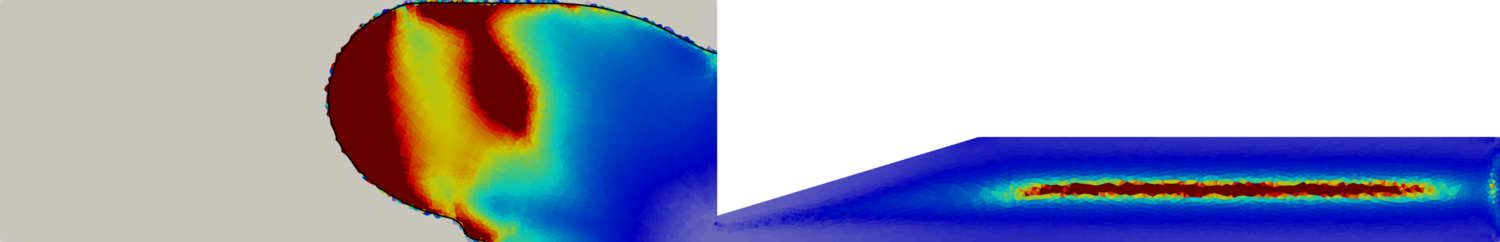}}
        \caption{FST (fine) at $t = \SI{0.1000}{s}$}
        \label{fig:2DSandiaCavityViscDt0_0001_40Nrec}
    \end{subfigure}
    ~ 
    \begin{subfigure}[t]{0.3\textwidth}
        \reflectbox{\includegraphics[width=\textwidth]{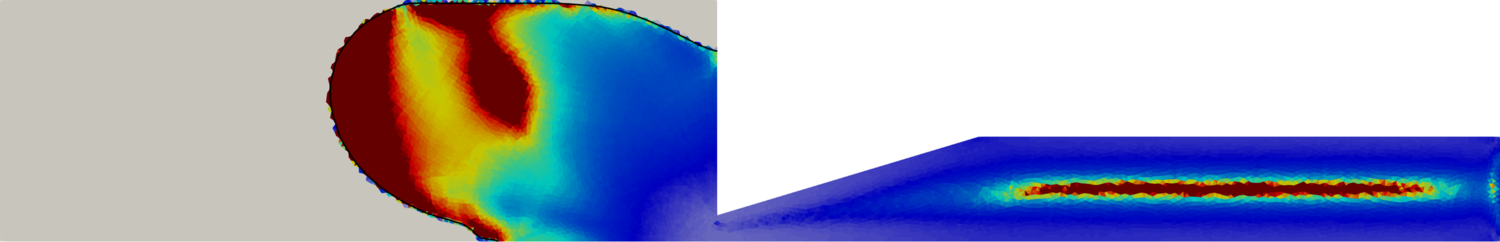}}
        \caption{SST at $t = \SI{0.1000}{s}$}
        \label{fig:2DSandiaCavity_ViscTimeRef40Nrec}
    \end{subfigure}
    ~ 
    \begin{subfigure}[t]{0.3\textwidth}
        \reflectbox{\includegraphics[width=\textwidth]{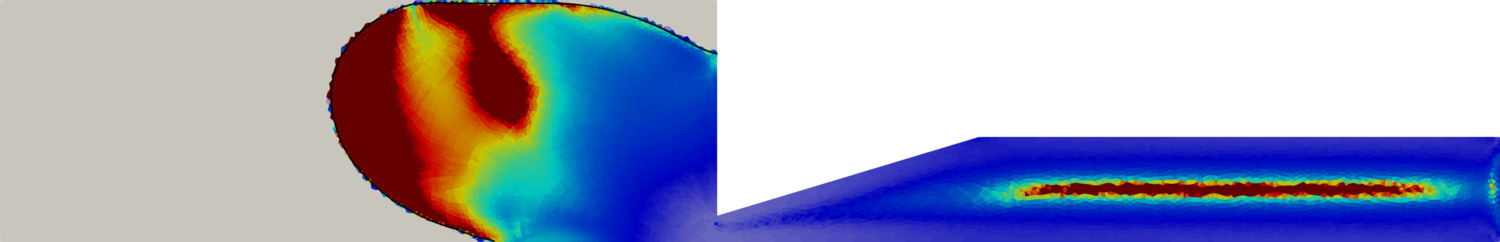}}
        \caption{FST (coarse) at $t = \SI{0.1000}{s}$}
        \label{fig:2DSandiaCavityViscDt0_0005_40Nrec}
    \end{subfigure}\\

    \caption{Molten material position at various time instances, obtained with the FST method of time-slab thickness, $\Delta t = \SI{0.0001}{s}$ (left column), the SST method with adaptive temporal refinement, $\Delta t = \SI{0.0001}{s} - \SI{0.0005}{s}$,  (middle column) and the FST method of time-slab thickness, $\Delta t = \SI{0.0005}{s}$ (right column).}
    \label{fig:StepCavity2DShape}
\end{figure}

The PZT ceramic paste fills the cavity for $\SI{0.1}{s}$. We use three different temporal discretizations here. The first one is a flat space-time (FST) discretization, consisting of $\SI{200}{}$ time slabs ($\Delta t = \SI{0.0005}{s}$), in contrast to an FST discretization of $\SI{1000}{}$ time slabs. Furthermore, we apply adaptive temporal refinement, as described in \cite{karyofylli2018simplex}, which leads to a hybrid simplex space-time (SST) discretization, with each of the $\SI{200}{}$ time slabs being $\SI{0.0005}{s}$ thick and discretized differently with $\SI{1}{}$ to $\SI{5}{}$ elements in the time direction (see Figure \ref{fig:TemporalRefinement}). Note that the temporal accuracy is increased only close to the propagating interface, as shown in Figure \ref{fig:TemporalRefinement}.

\begin{figure}[!htb]
    \centering
    \begin{subfigure}[t]{0.75\textwidth}
        \includegraphics[width=\textwidth]{./Pictures/2DSandiaCavity/ViscosityRamp.png}
        \label{fig:2DSandiaCavityVelocityRam}
    \end{subfigure}\\
    
    \begin{subfigure}[t]{0.3\textwidth}
        \centering
        \includegraphics[width=1.0\textwidth]{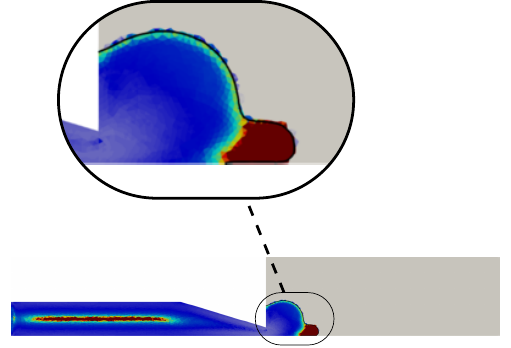}
        \caption{FST (fine)}
        \label{fig:Visc0_0001_22Nrec}
    \end{subfigure}
    ~
    \begin{subfigure}[t]{0.3\textwidth}
        \centering
        \includegraphics[width=1.0\textwidth]{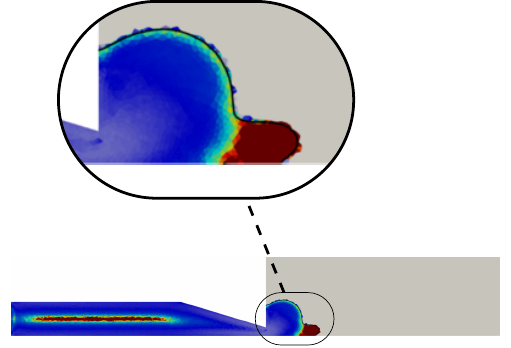}
        \caption{SST}
        \label{fig:ViscTRef22Nrec}
    \end{subfigure}
    ~
    \begin{subfigure}[t]{0.3\textwidth}
        \centering
        \includegraphics[width=1.0\textwidth]{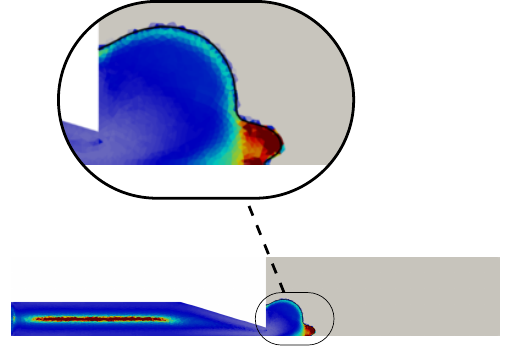}
        \caption{FST (coarse)}
        \label{fig:Visc0_0005_22Nrec}
    \end{subfigure}
    \caption{Detailed comparison of the interface topology and viscosity computed by different space-time formulations at $t = \SI{0.0550}{s}$.}
    \label{fig:Sandia2DFountain}
\end{figure}

Figure \ref{fig:StepCavity2DShape} illustrates the front position of the molten material at various time instances. The results obtained with the fine and coarse FST discretization are then compared with those computed with the SST discretization while applying temporal refinement. As we can realize from Figure \ref{fig:Sandia2DFountain}, the wetting, and the viscosity profile are better resolved by the SST and the finer FST discretization compared to the coarse FST discretization because they provide higher temporal accuracy.

\begin{figure}[!h]
    \begin{minipage}{0.525\textwidth}
        \vspace{+0.65cm}
        \begin{subfigure}{0.1\textwidth}
            \centering
            \raisebox{+0.5cm}{\includegraphics[width=0.1\textwidth]{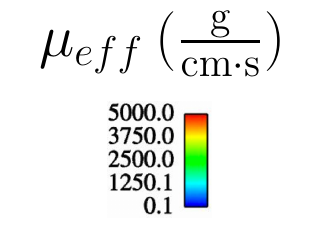}}
        \end{subfigure}
        ~
        \begin{subfigure}{0.8\textwidth}
            \centering
            \includegraphics[width=\textwidth]{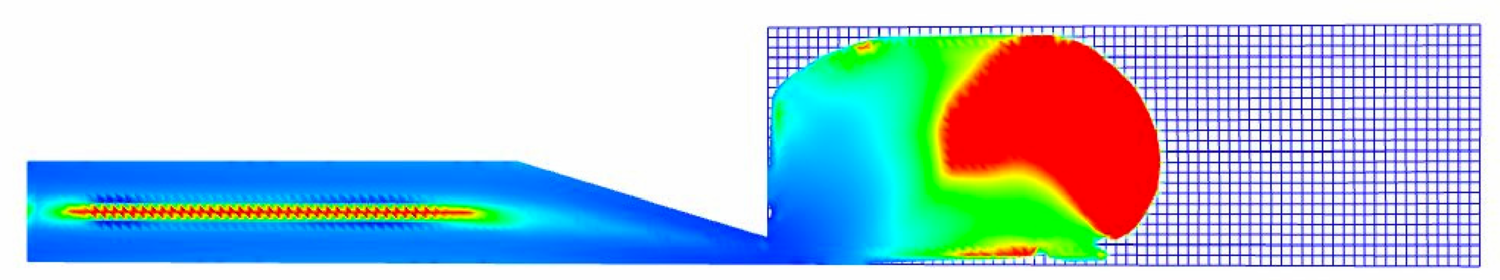}
            \caption{Sandia National Laboratories}
            \label{SNL}
        \end{subfigure}
    \end{minipage}
    \hfill
    \begin{minipage}{0.425\textwidth}
        \begin{subfigure}{\textwidth}
            \centering
            \includegraphics[width=\textwidth]{Pictures/2DSandiaCavity/ViscosityRamp}
        \end{subfigure}
        \par\medskip
        \begin{subfigure}{\textwidth}
            \centering
            \reflectbox{ \includegraphics[width=\textwidth]{Pictures/2DSandiaCavity/ViscDt0_0001_40Nrec}}
            \caption{FST (fine)}
            \label{XNS}
        \end{subfigure}
    \end{minipage}%
\caption{Comparison between the viscosity profile of a non-isothermal Carreau-WLF fluid, as published by \cite{Sandia} (\ref{SNL}) and that obtained with our in-house solver (\ref{XNS}), at the same filling stage.}
\label{fig:SandiaNonIsothermalViscComparison}
\end{figure}

Furthermore, as Figure \ref{fig:SandiaNonIsothermalViscComparison} indicates, the simulation with the FST discretization leads to a viscosity distribution qualitatively similar to that presented in \cite{Sandia}. Inside the pipe, the viscosity remains high in the center and low near the walls, whereas inside the cavity, the viscosity increases progressively. This especially holds close to the interface, in both simulations. It is worth mentioning that a linearized Blake model in combination with a Navier-slip boundary condition regarding the velocity degrees of freedom is used in \cite{Sandia}. Furthermore, in contrast to the results presented here, a heat flux boundary condition was applied to the walls of the steel mold. Another difference is that, in this work, a finer spatial discretization for the computations is used. All the factors mentioned above could explain the quantitative differences, which exist between the two results.

In Table \ref{table:Performance2D}, we present the performance of the simulations, using three different temporal discretizations, as described above. All simulations were performed in parallel (MPI paralleliztion), utilizing \(\SI{64}{}\) cores on the RWTH Aachen University IT Center cluster.
\begin{table}[ht]
\centering
\captionsetup{justification=centering}
\caption{Typical computational performance of 2D mold filling computations.}
\resizebox{0.95\textwidth}{!}{\tabcolsep7pt
\begin{tabular}{c c c c c c}
\toprule
 & \bf{Time}   & \bf{Nodes}    & \bf{Elements}  & \bf{Total time for}                        & \bf{Total time for} \\
 & \bf{Steps} & \bf{per step}  & \bf{per step}   & \bf{system formation (\si{s})}       & \bf{system solution (\si{s})}   \\ [0.5ex] 
\midrule
FST (fine)            & \(\SI{1000}{}\) &         \(\SI{28024}{}\)   &          \(\SI{27334}{}\)   &  \(\SI{555.09}{}\)   & \(\SI{16186.13}{}\)   \\
FST (coarse)        &   \(\SI{200}{}\) &          \(\SI{28024}{}\)  &          \(\SI{27334}{}\)   &  \(\SI{76.61}{}\) & \(\SI{4100.53}{}\) \\
SST                     &   \(\SI{200}{}\) & \(\SI{\sim 31872}{}\) & \(\SI{\sim 104546}{}\) &  \(\SI{86.42}{}\)   & \(\SI{4351.73}{}\)   \\ [1ex]
\bottomrule
\end{tabular}}
\label{table:Performance2D}
\end{table}

We perform Newton-Raphson iterations at every time step, because of the nonlinearity of the Navier-Stokes equations and the heat equation, due to the viscous dissipation term. Strong coupling iterations are also executed due to the mutual dependence between the Navier-Stokes, the level-set, and the heat equation. A GMRES solver is employed for solving the resulting linear systems of equations, in combination with an ILUT factorization. To summarize, the use of SST discretization combined with adaptive temporal refinement leads to a similar resolution of the space-time evolving interface and the viscosity distribution as the one achieved by the fine FST discretization, while keeping the total time for forming and solving the system around $70 \%$ lower than in the case of the fine FST simulation (cf. Fig. \ref{fig:Sandia2DFountain}). 

\subsection{3D cavity with a distributor}
As a second test case, we compute the filling process of a 3D coat hanger distributor and a cavity. The inflow is at the small end of the pipe, whereas the outflow is at the back of the cavity on the opposite side. Thanks to the geometry's symmetry relative to the center plane, only half of the geometry is simulated, saving time and resources. The same test case was simulated in \cite{Sandia} with the software ARIA.
\begin{figure}[!htb]
    \centering
    \includegraphics[width=0.75\textwidth]{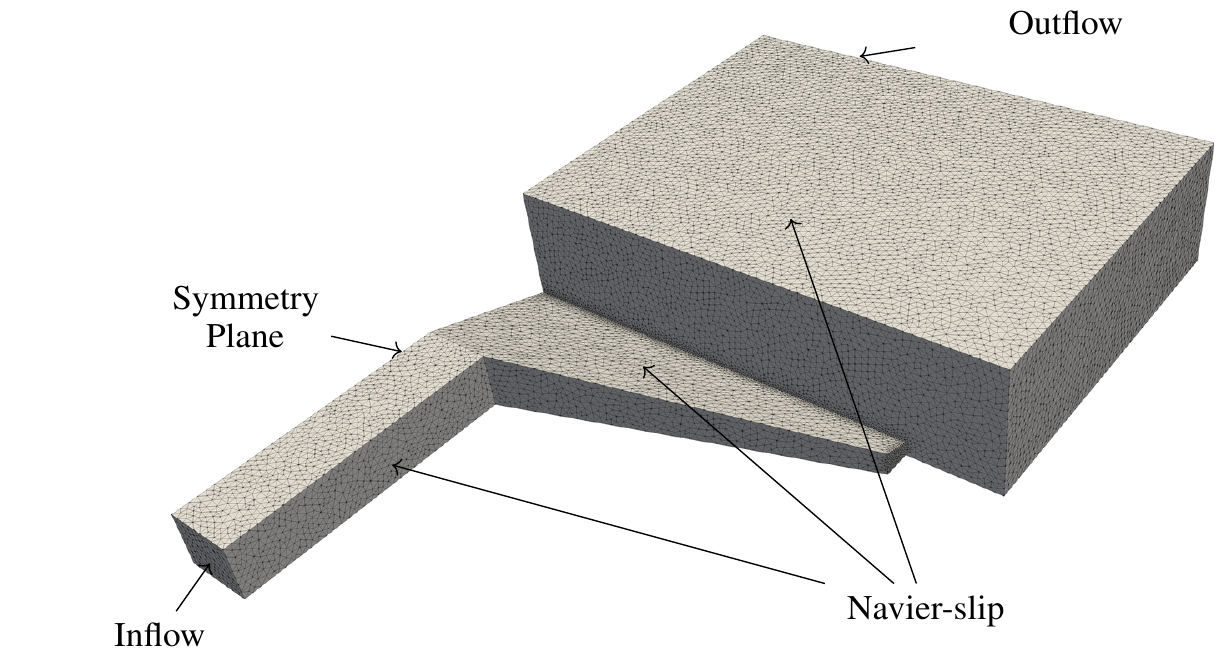}
    \caption{The computational domain of the three dimensional cavity with a distributor and a pipe, as described by Rao et al. \cite{Sandia}. Only half of the geometry is depicted due to symmetry.} \label{fig:Sandia3D}
\end{figure}

In this benchmark, we observe the temperature and viscosity profile, as well as the shape of the evolving interface. Although a Newtonian melt without any temperature gradients is assumed in \cite{Sandia}, shear-thinning effects are taken into consideration in our first computation and represented with the Carreau-WLF model. The domain is illustrated in Figure \ref{fig:Sandia3D}. For our first simulation, we use a very fine spatial discretization, consisting of \(\SI{7119241}{}\) tetrahedral elements and the time-slab thickness is \(\Delta t = \SI{0.0005}{s}\). Navier-slip boundary condition is assumed on the walls for the velocity degrees of freedom, except for the inflow, outflow and symmetry plane. The Blake wetting condition is not taken into consideration here.  A constant parabolic velocity profile, which leads to an average velocity of \(\SI{190}{\frac{cm}{s}}\), is imposed at the inflow boundary, whereas traction-free boundary conditions are used at the outflow boundary. A slip boundary condition is used for the symmetry plane. Concerning the temperature degree of freedom, we impose a Dirichlet boundary condition on the mold walls \(T = \SI{298}{K}\), the injection temperature is equal to \(\SI{405}{K}\) at the inflow, and we assume homogeneous Neumann boundary conditions at the symmetry plane. Although we consider a non-isothermal condition, the phase-change effects are disregarded. The material properties are given in Table \ref{table:MaterialProperties2D}. The gravitational acceleration is equal to \(f_y  = -g = \SI{-981}{\frac{cm}{s^2}}\) and the surface tension coefficient \(\gamma\) is set to \(\SI{42.4}{\frac{g}{s^2}}\).

\begin{figure}[!h]
    \begin{minipage}{0.245\textwidth}
        \begin{subfigure}{\textwidth}
            \centering
            \includegraphics[width=0.90\textwidth]{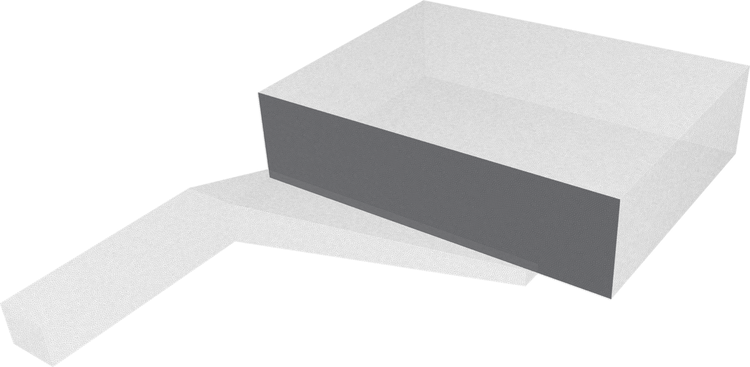}
            \caption{Slice Position}
            \label{SlicePos}
        \end{subfigure}
    \end{minipage}%
    \hfill
    \begin{minipage}{0.245\textwidth}
        \begin{subfigure}{\textwidth}
            \centering
            \includegraphics[width=\textwidth]{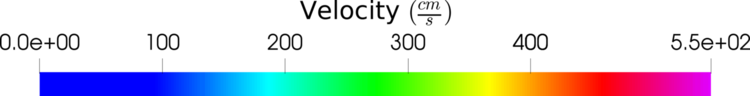}
        \end{subfigure}
        \par\medskip
        \begin{subfigure}{\textwidth}
            \centering
            \includegraphics[width=\textwidth]{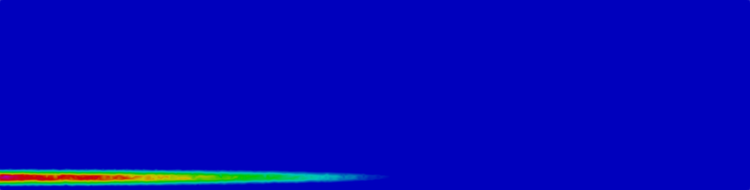}
            \caption{Velocity}
            \label{VelProf}
        \end{subfigure}
    \end{minipage}%
    \hfill
    \begin{minipage}{0.245\textwidth}
        \begin{subfigure}{\textwidth}
            \centering
            \includegraphics[width=\textwidth]{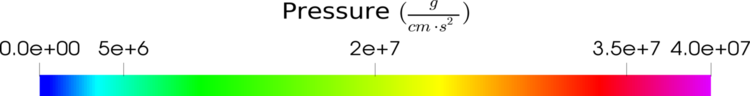}
        \end{subfigure}
        \par\medskip
        \begin{subfigure}{\textwidth}
            \centering
            \includegraphics[width=\textwidth]{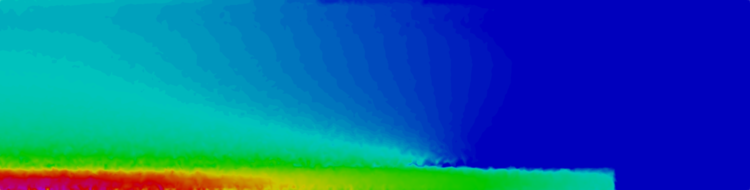}
            \caption{Pressure}
            \label{PresProf}
        \end{subfigure}
    \end{minipage}%
    \hfill
    \begin{minipage}{0.245\textwidth}
        \begin{subfigure}{\textwidth}
            \centering
            \includegraphics[width=\textwidth]{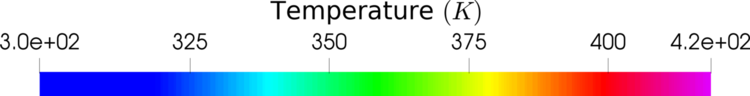}
        \end{subfigure}
        \par\medskip
        \begin{subfigure}{\textwidth}
            \centering
            \includegraphics[width=\textwidth]{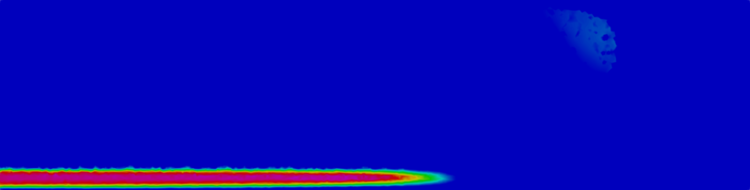}
            \caption{Temperature}
            \label{TempProf}
        \end{subfigure}
    \end{minipage}%
\caption{Position of a slice plane close to the inflow of the 3D cavity (\ref{SlicePos}) and the velocity (\ref{VelProf}), pressure (\ref{PresProf}) and temperature field (\ref{TempProf}) along this plane at $t=\SI{0.125}{s}$.}
\label{fig:Sandia3DVelPresTemp}
\end{figure}

In Figure \ref{fig:Sandia3DVelPresTemp}, we show the velocity, pressure and temperature fields on a slice plane inside the cavity close to the inflow from the distributor at $t=\SI{0.125}{s}$. As expected, the highest velocity occurs at the inflow of the cavity, whereas at the walls the velocity remains close to zero due to the Navier-slip boundary condition. The temperature distribution exhibits similar behavior since we assume adiabatic walls. We observe a gradient of pressure at the position, where the distributor opens to the mold, with the highest pressure happening at the center of the channel.

\begin{figure}[!h]
    \begin{minipage}{0.475\textwidth}
        \vspace{+1.15cm}
        \begin{subfigure}{1.0\textwidth}
            \centering
            \includegraphics[width=\textwidth]{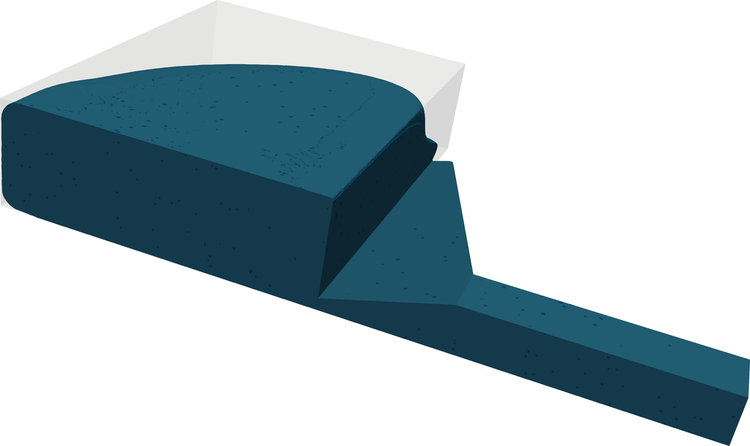}
            \caption{Surface profile of the melt}
            \label{MeltSurf}
        \end{subfigure}
    \end{minipage}
    \hfill
    \begin{minipage}{0.475\textwidth}
        \begin{subfigure}{\textwidth}
            \centering
            \includegraphics[width=\textwidth]{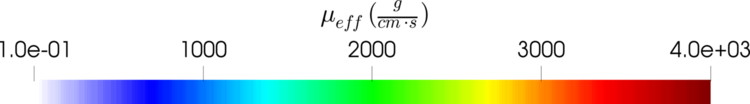}
        \end{subfigure}
        \par\medskip
        \begin{subfigure}{\textwidth}
            \centering
            \includegraphics[width=\textwidth]{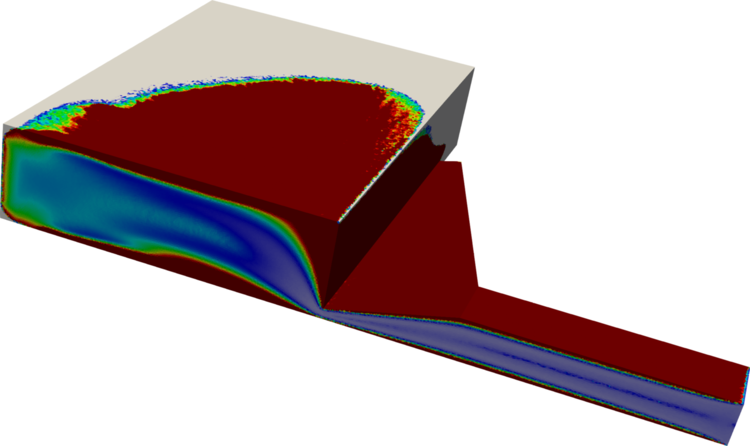}
            \caption{Viscosity profile}
            \label{ViscProf}
        \end{subfigure}
    \end{minipage}%
\caption{Shape of the melt-air interface and the viscosity distribution at $t=\SI{0.09}{s}$.}
\label{fig:Sandia3DSurf&ViscProf}
\end{figure}

Figure \ref{fig:Sandia3DSurf&ViscProf} shows the shape of the front between the molten PZT and air as well as the viscosity field at $t=\SI{0.09}{s}$. As seen, the viscosity values close to the walls of the mold are over \(\SI{4000}{\frac{g}{cm\cdot s}}\). This observation is of great significance because the viscosity values on the cavity boundaries influence the wetting behavior of the melt the most, as already noted in \cite{Sandia}. As a consequence, we decided to perfom the same filling simulation, considering this time isothermal conditions and the PZT ceramic paste as Newtonian fluid with constant viscosity of value \SI{4000}{\frac{g}{cm \cdot s}}. For maintaining a good convergence rate, we used the following properties for air: $\rho_{air} = \SI{0.0045}{\frac{g}{cm^3}}$ and $(\mu_{\mathit{eff}})_{air} = \SI{4.0}{\frac{g}{cm \cdot s}}$.

\begin{figure}[!htb]
    \centering
    \begin{subfigure}[t]{0.24\textwidth}
        \centering
        \includegraphics[scale=1.0]{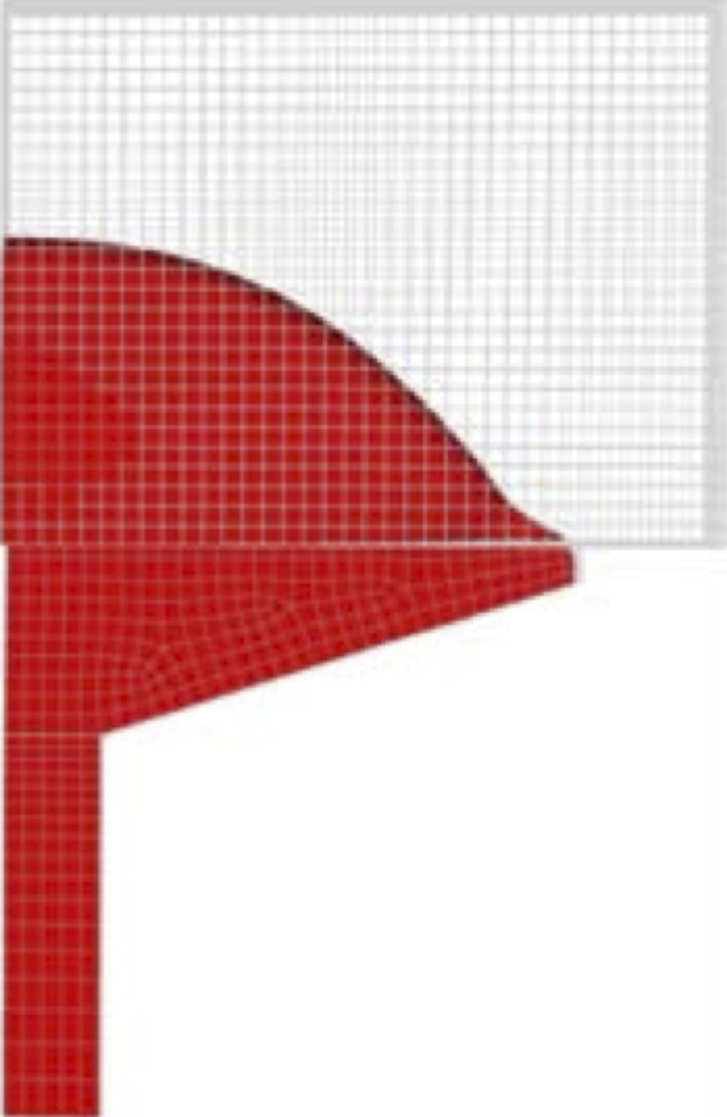}
        \label{fig:SandiaRef3D05s}
    \end{subfigure}
    \hfill  
    \begin{subfigure}[t]{0.24\textwidth}
        \centering
        \includegraphics[width=0.75\textwidth]{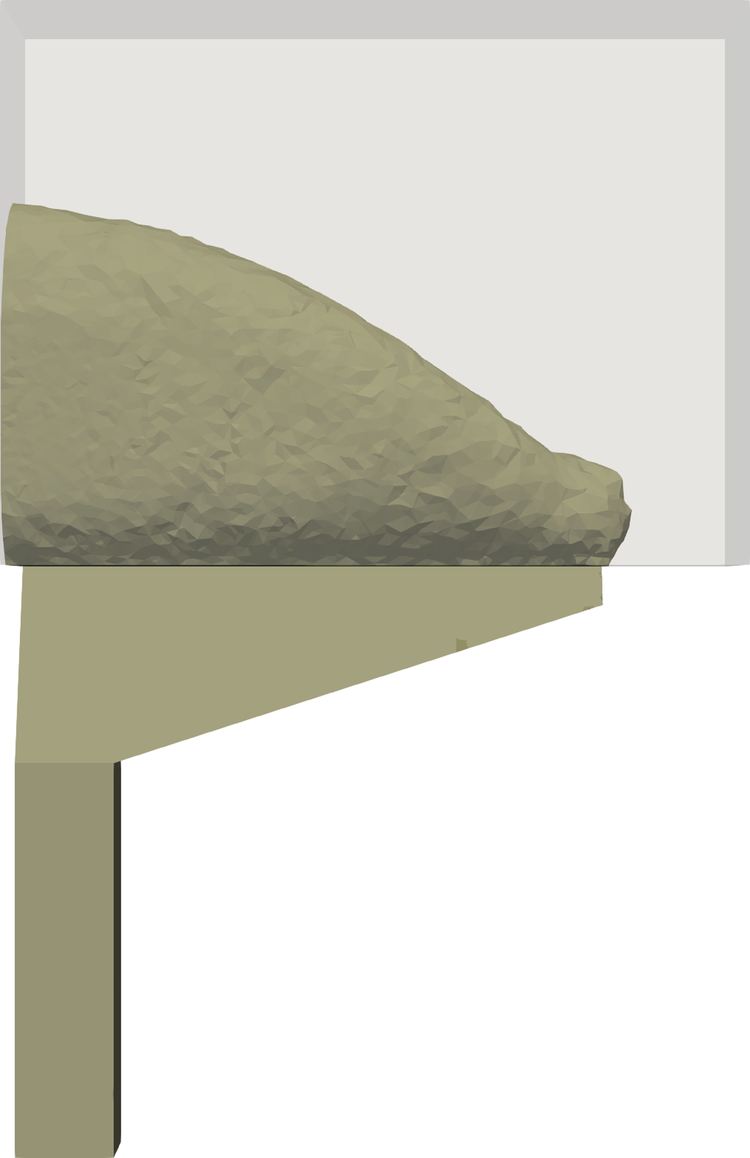}
        \label{fig:FSTFine05}
    \end{subfigure}
    \hfill  
    \begin{subfigure}[t]{0.24\textwidth}
        \centering
        \includegraphics[width=0.75\textwidth]{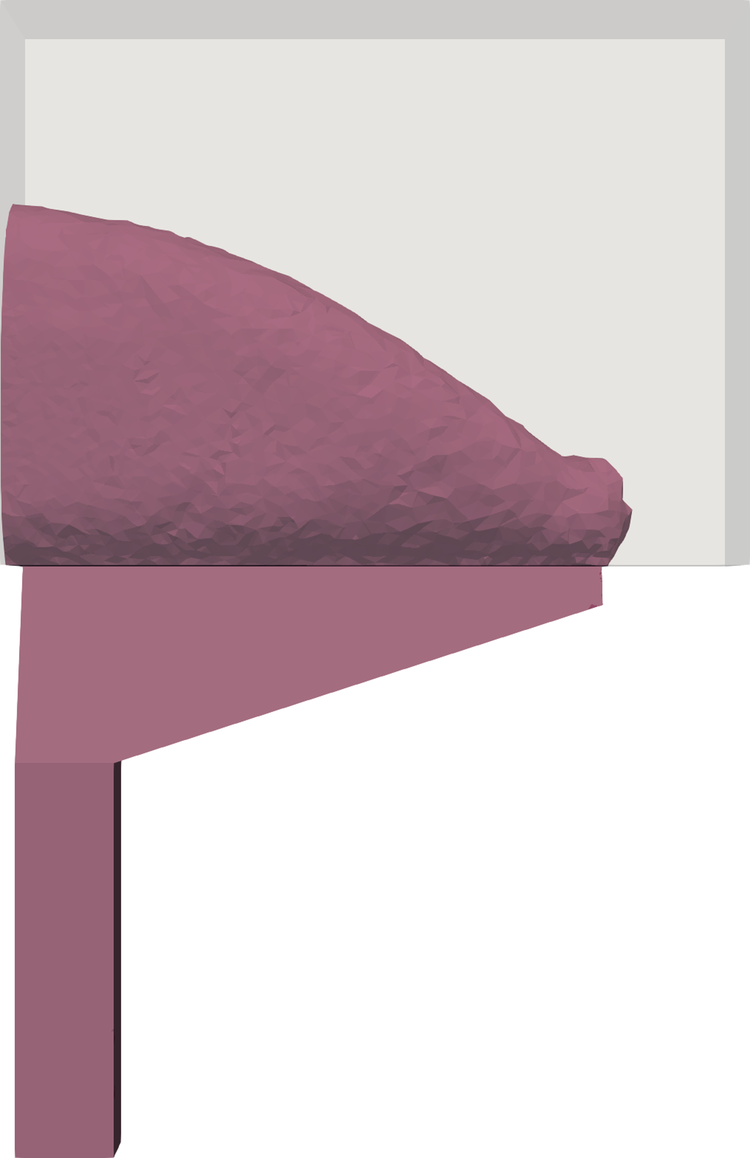}
        \label{fig:SST05}
    \end{subfigure}
    \hfill  
    \begin{subfigure}[t]{0.24\textwidth}
        \centering
        \includegraphics[width=0.75\textwidth]{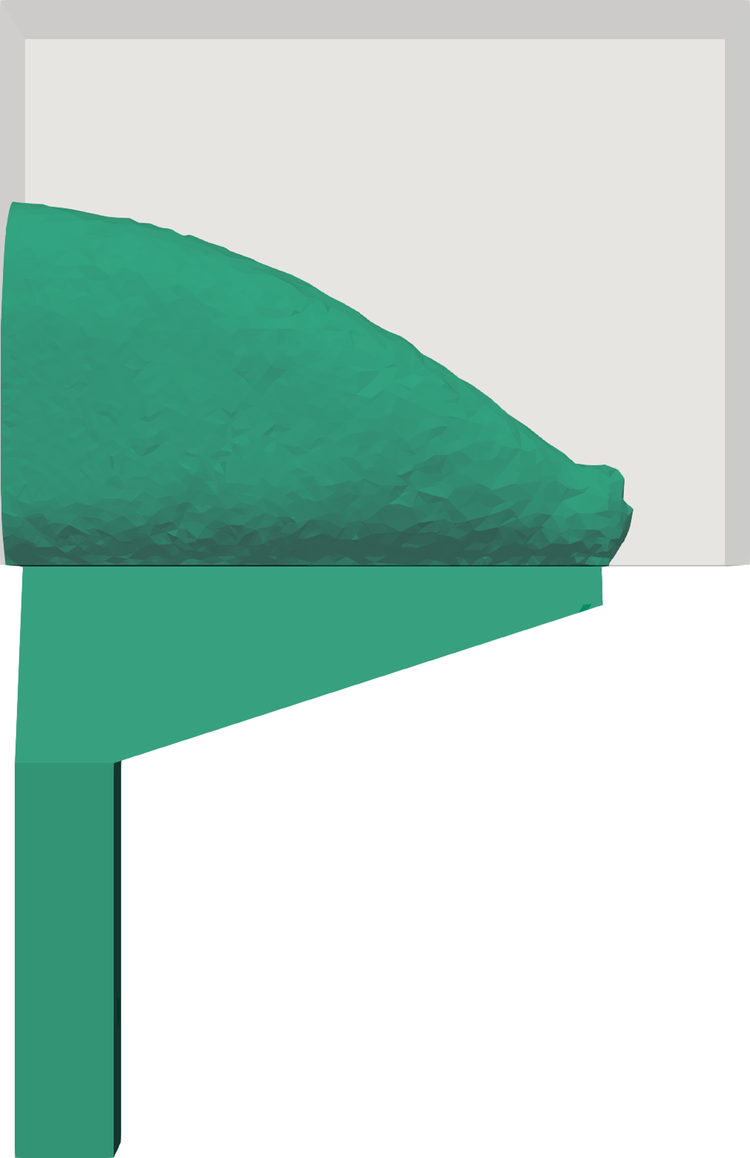}
        \label{fig:FSTCoarse05}
    \end{subfigure}
    \par\medskip 
    \begin{subfigure}[t]{0.24\textwidth}
        \centering
        \includegraphics[scale=1.0]{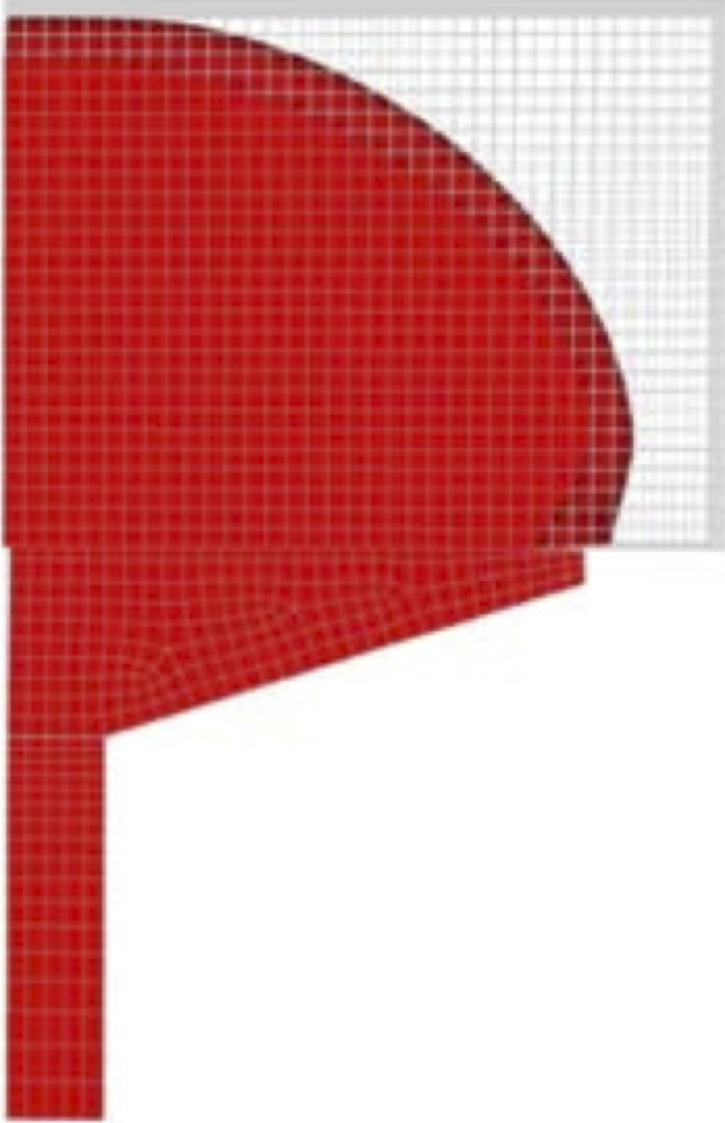}
        \caption{Sandia National Laboratories}
        \label{fig:SandiaRef3D09s}
    \end{subfigure}
    \hfill  
    \begin{subfigure}[t]{0.24\textwidth}
        \centering
        \includegraphics[width=0.75\textwidth]{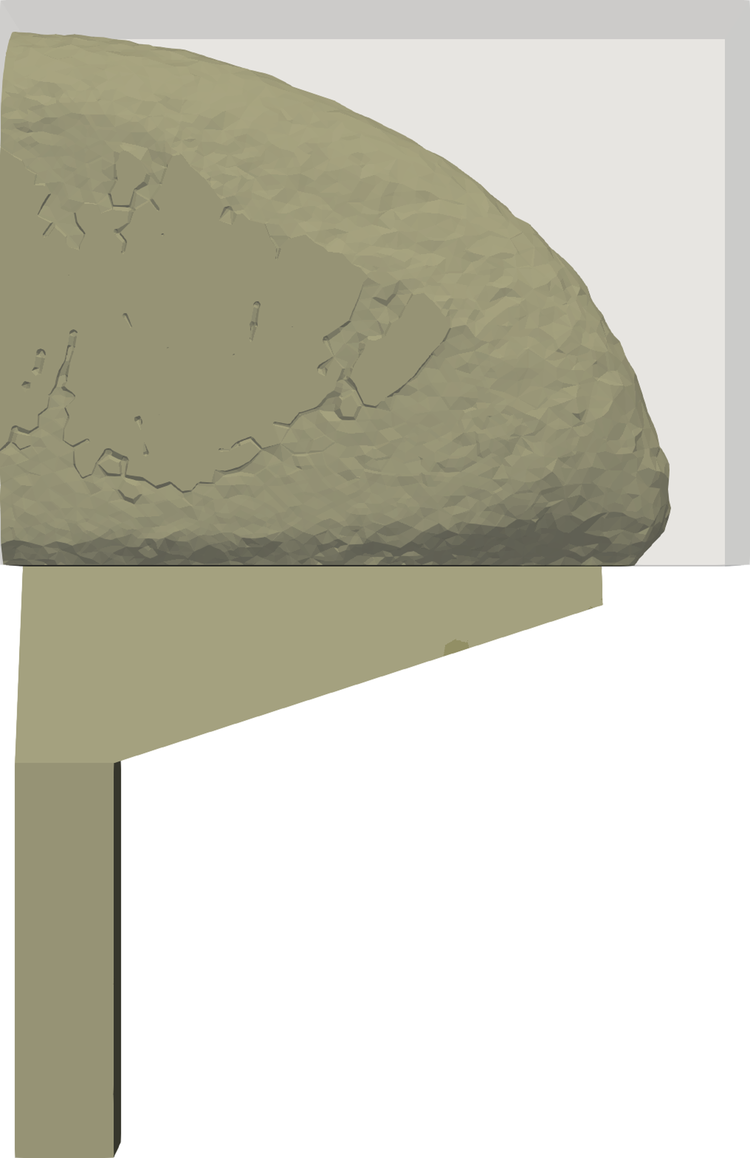}
        \caption{FST (fine)}
        \label{fig:FSTFine09}
    \end{subfigure}
    \hfill  
    \begin{subfigure}[t]{0.24\textwidth}
        \centering
        \includegraphics[width=0.75\textwidth]{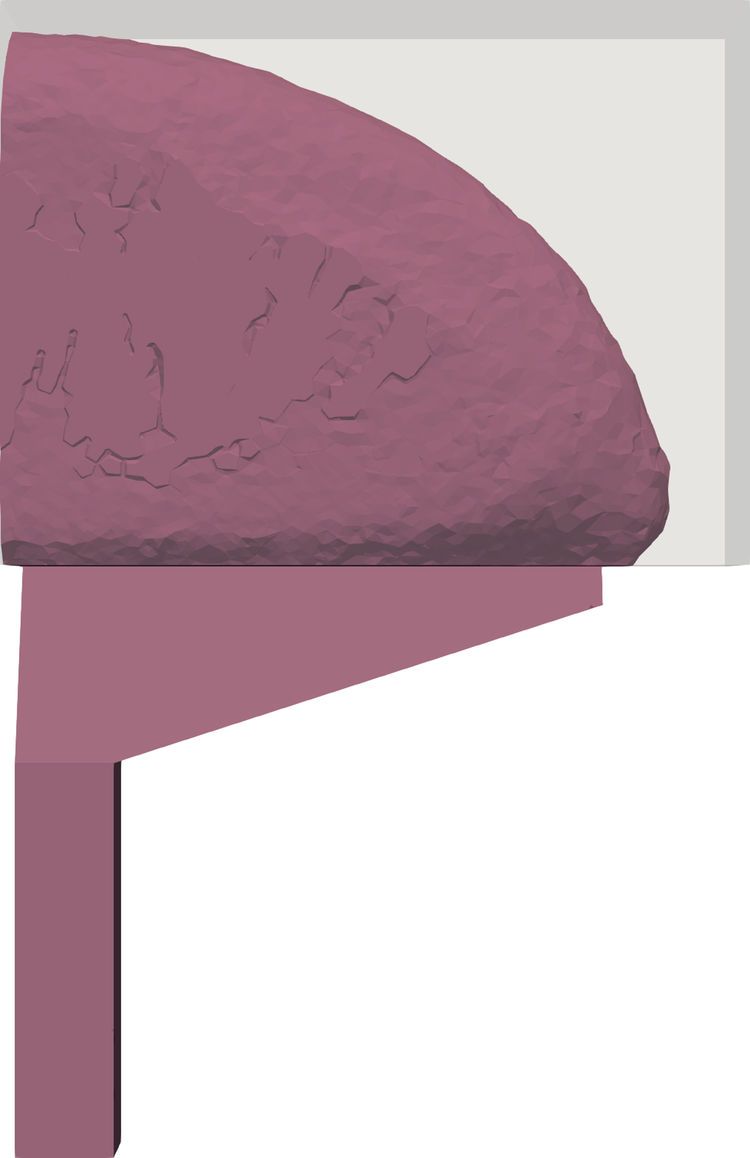}
        \caption{SST}
        \label{fig:SST09}
    \end{subfigure}
    \hfill  
    \begin{subfigure}[t]{0.24\textwidth}
        \centering
        \includegraphics[width=0.75\textwidth]{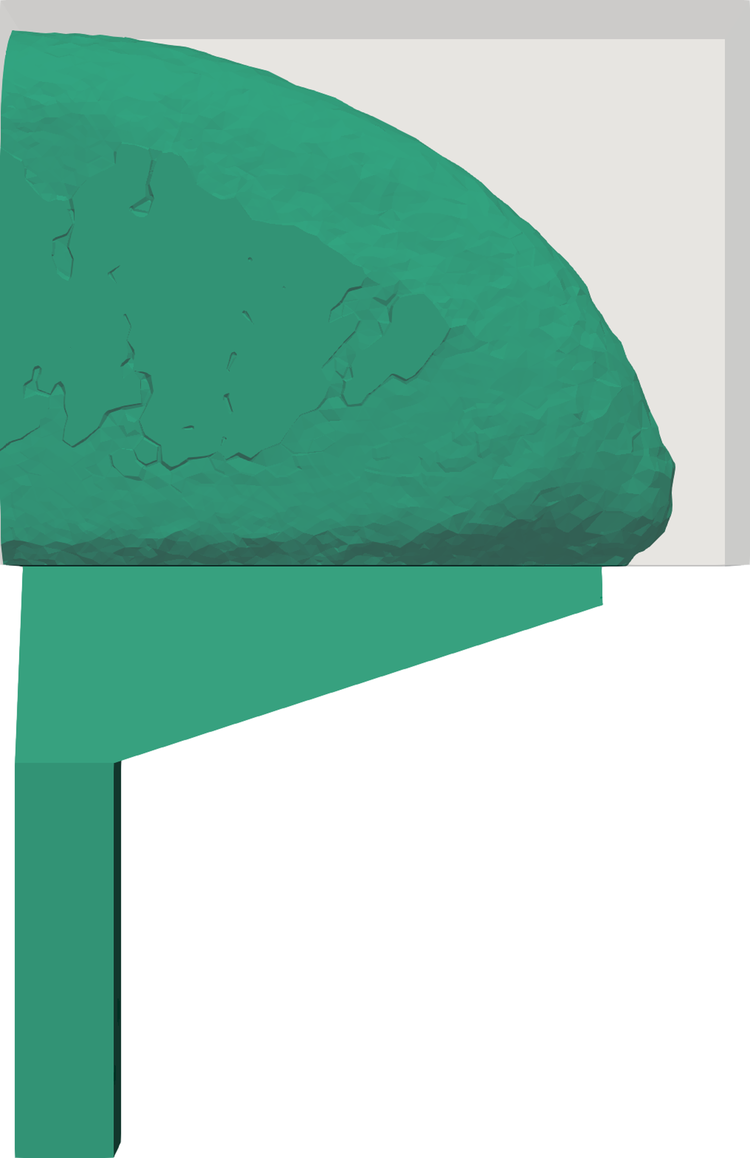}
        \caption{FST (coarse)}
        \label{fig:FSTCoarse09}
    \end{subfigure}
    \caption{Molten material position at \(t= \SI{0.5}{s}\) (top row) and \(t= \SI{0.9}{s}\) (bottom row), obtained with our in-house solver (right columns) and compared with reference data (leftmost column) from \cite{Sandia}.}
    \label{fig:Sandia3DShapeComparison}
\end{figure}

For the later simulations, the spatial discretization is much coarser and consists of \(\SI{176398}{}\) tetrahedral elements. We made use of three different temporal discretizations, which are the same as in the 2D benchmark case. In Figure \ref{fig:Sandia3DShapeComparison}, we compare the molten material interface at two different time instances, as obtained with our in-house solver utilizing three different temporal discretizations, and the one shown in reference data \cite{Sandia}. In Table \ref{table:Performance3D}, we present the timings of our 3D simulations, employing three different temporal discretizations. All three simulations were computed again in parallel (MPI parallelization) utilizing \(\SI{64}{}\) cores on the RWTH Aachen University IT Center cluster. Similar to the 2D filling case, the use of SST discretization combined with adaptive temporal refinement reduces the total time for forming and solving the system by roughly $55 \%$, compared to the case of the fine FST simulation.
\begin{table}[!htb]
\centering
\captionsetup{justification=centering}
\caption{Typical performance of 3D mold filling computations.}
\resizebox{0.95\textwidth}{!}{\tabcolsep7pt
\begin{tabular}{c c c c c c}
\toprule
 & \bf{Time}   & \bf{Nodes}    & \bf{Elements}  & \bf{Total time for}                        & \bf{Total time for} \\
 & \bf{Steps} & \bf{per step}  & \bf{per step}   & \bf{system formation (\si{s})}       & \bf{system solution (\si{s})}   \\ [0.5ex] 
\midrule
FST (fine)            & \(\SI{1000}{}\) &         \(\SI{70542}{}\)   &          \(\SI{176398}{}\)   &  \(\SI{9298.96}{}\)   & \(\SI{7534.43}{}\)   \\
FST (coarse)        &   \(\SI{200}{}\) &          \(\SI{70542}{}\)  &          \(\SI{176398}{}\)   &  \(\SI{1923.78}{}\) & \(\SI{1666.57}{}\) \\
SST                     &   \(\SI{200}{}\) & \(\SI{\sim 100790}{}\) & \(\SI{\sim 1325318}{}\) &  \(\SI{2624.87}{}\)   & \(\SI{3307.29}{}\)   \\ [1ex]
\bottomrule
\end{tabular}}
\label{table:Performance3D}
\end{table}

\section{Conclusions}
In this paper, the non-isothermal two-phase flow of a highly viscous fluid has been computed and shows shear-thinning effects, during the filling process of injection molding. The Carreau-WLF material model described the behavior of the melt. For achieving a better wetting on the mold walls, the Navier-slip boundary condition has been used.

Besides, a novel discretization approach has been presented, which allows arbitrary temporal refinement of the space-time slabs in the vicinity of the evolving front of the highly viscous molten material during injection molding. Future work includes the combination of arbitrary temporal refinement with arbitrary spatial refinement close to the propagating interface. The refinement criterion should be based on an appropriate a-posteriori error estimate of local or global quantities of interest, such as the front curvature, pressure jumps close to the interface, material discontinuities and the gradient in the viscosity field.

\section*{Acknowledgment}
The authors gratefully acknowledge the support of the German Research Foundation (DFG) under program SFB 1120 ``Precision Melt Engineering". The computations were conducted on computing clusters provided by the RWTH Aachen University IT Center and by the J\"ulich Aachen Research Alliance (JARA). Furthermore, we would like to thank the student Efstratios Moskofidis for his contribution by simulating different benchmark cases and our colleague, Max von Danwitz for the fruitful discussions.


\end{document}